\documentclass[12pt,preprintnumbers]{article}
\usepackage{amsmath, amssymb, wasysym, slashed, multirow,hyperref}
\usepackage{pdfpages}
\usepackage{cite}
\usepackage{rotating}
\usepackage{times}

\newenvironment{sciabstract}{%
\begin{quote} }
{\end{quote}}

\newcounter{lastnote}

\usepackage{fancyhdr}
\fancypagestyle{plain}{%
	\fancyhead[R]{ACFI-T18-21}
	
}

\begin{document}
\title{Probing Neutrino Dirac Mass in Left-Right Symmetric Models at the LHC and Next Generation Colliders }

% Place the author information here.  Please hand-code the contact
% information and notecalls; do *not* use \footnote commands.  Let the
% author contact information appear immediately below the author names
% as shown.  We would also prefer that you don't change the type-size
% settings shown here.

\author
{Juan Carlos Helo$^{1\ast}$, Haolin Li$^{2,6\dagger}$, Nicol\'as A. Neill$^{3,4\ddag}$,\\ Michael Ramsey-Musolf$^{2,5\S}$ and
  Juan Carlos Vasquez$^{3,4\circledast}$ \\
Departamento de F\'isica, Facultad de Ciencias, Universidad de La Serena, \\
 La Serena, Chile$^{1}$\\
 Amherst Center for Fundamental Interactions, Department of Physics, \\ University of Massachusetts-Amherst, U.S.A$^{2}$\\
Departamento de F\`isica Universidad T\'ecnica Federico Santa Mar\'ia, \\ Valpara\'iso, Chile$^{3}$\\
Centro Cientifico Tecnologico de Valparaiso, Valparaiso, Chile$^{4}$\\
Kellogg Radiation Laboratory, California Institute of Technology,\\ Pasadena, CA 91125 USA$^{5}$\\
Institute of Theoretical Physics, China Academic of Science, \\Beijing 100190, China$^{6}$\\
\small{ E-mail: jchelo@userena.cl$^*$, lihaolin1991@gmail.com$^{\dagger}$, mjrm@physics.umass.edu$^{\S}$,}\\
\small{nicolas.neill@gmail.com$^{\ddag}$, juan.vasquezcar@usm.cl$^{\circledast}$}
}

% Include the date command, but leave its argument blank.

\date{}

%%%%%%%%%%%%%%%%% END OF PREAMBLE %%%%%%%%%%%%%%%%


\baselineskip16pt


\maketitle


\begin{sciabstract}
We assess the sensitivity of the LHC, its high energy upgrade, and a prospective $100$ TeV hadronic collider to the Dirac Yukawa coupling of the heavy neutrinos in left-right symmetric models (LRSMs). We focus specifically on the trilepton final state in regions of parameter space yielding prompt decays of the right-handed gauge bosons ($W_R$) and neutrinos ($N_R$). In the minimal LRSM, the Dirac Yukawa couplings are completely fixed in terms of the mass matrices for the heavy and light neutrinos. In this case, the trilepton signal provides a direct probe of the Dirac mass term for a fixed $W_R$ and $N_R$ mass. 
We find that while it is possible to discover the $W_R$ at the LHC, probing the Dirac Yukawa couplings will require a 100 TeV $pp$ collider. We also show that the observation of the trilepton signal at the LHC would indicate the presence of a non-minimal LRSM scenario. %there still some room for finding the $W_R$ boson at the LHC, while the Dirac Yukawa coupling for heavy neutrino could be probed at the 100 TeV collider.         
\end{sciabstract}

% In setting up this template for *Science* papers, we've used both
% the \section* command and the \paragraph* command for topical
% divisions.  Which you use will, of course, depend on the type of paper
% you're writing.  Review Articles tend to have displayed headings, for
% which \section* is more appropriate; Research Articles, when they have
% formal topical divisions at all, tend to signal them with bold text
% that runs into the paragraph, for which \paragraph* is the right
% choice.  Either way, use the asterisk (*) modifier, as shown, to

\section{Introduction}

Soon after the appearance of the original works~\cite{Pati:1974yy,Mohapatra:1974gc,Senjanovic:1975rk}, the minimal left-right symmetric model (mLRSM)  has been proposed to connect the smallness of neutrino masses with the spontaneous violation of parity~\cite{Senjanovic:1978ev,Mohapatra:1979ia,Mohapatra:1980yp}. The origin of neutrino masses within the mLRSM must be understood in analogy with the explanation of the origin of mass within the Standard Model (SM). In the SM, fermion masses are obtained through the Higgs mechanism, for which one manifestation is the proportionality of a given Higgs boson fermionic branching ratio to the square of the corresponding fermion mass.
The Higgs boson has been discovered at the LHC by the ATLAS and CMS collaborations~\cite{Aad:2015zhl}, and the measured branching ratio into bottom quark and $\tau$ lepton pairs agree with the SM expectations~\cite{Chatrchyan:2014vua}.

In the neutrino sector, the situation becomes less clear since neutrinos are electrically neutral. While an SM-like pure Dirac mass is a possibility, the magnitudes of associated Yukawa couplings would be considerably smaller than for the charged elementary fermions. A theoretically attractive alternative is the see-saw mechanism~\cite{Minkowski:1977sc,Mohapatra:1979ia,Glashow:1979nm,GellMann:1980vs,Yanagida:1979as,Schechter:1980gr}, which exploits the possibility that the electrically neutral neutrino may be its own antiparticle. Neutrino masses in the mLRSM arise from a combination of two versions of the see-saw mechanism, the so-called Type-I, and Type-II variants.  

If left-right symmetry is realized in nature, it will be important to establish whether the mLRSM is at the same level as the SM regarding the origin of fermion masses.  Soon after its original proposal, approaches for probing the Yukawa couplings of heavy and light neutrinos were considered. {The Yukawa sector for the light neutrinos may be in principle probed in low energy experiments, such as neutrinoless double beta decay and oscillation experiments}. The Yukawa couplings of heavy neutrinos (HNs) can be probed in high energy experiments through the Keung-Senjanovi\'c (KS)  process~\cite{Keung:1983uu} , which consists in the production of an on-shell, heavy $W_R$ gauge boson decaying into two right-handed leptons and two jets. { More recently, in Refs.~\cite{Nemevsek:2012iq,Senjanovic:2016vxw,Senjanovic:2018xtu} it was found that  the Dirac Yukawa coupling of neutrinos (which is proportional to the mixing between heavy and light neutrinos) can be unambiguously obtained  once all light and heavy neutrino masses and mixing angles are measured.} Therefore, this puts the mLRSM as a \emph{testable} model of neutrino masses and calls for the experimental verification of the relation between the Dirac mass and the heavy and light neutrino masses,  the main subject of this work. %\textcolor{blue}{Notice that this situation is in complete analogy with the SM in the sense that, once the Higgs vacuum expectation value and the quark masses are known the quark Yukawa couplings are uniquely fixed. }

It is worth emphasizing that without left-right symmetry, the connection between the Dirac mass matrix and the heavy and light neutrino mass matrices is lost. This can be explicitly seen in the Casas-Ibarra parametrization~\cite{Casas:2001sr}, where the Dirac mass matrix is given in terms of the heavy and light neutrino mass matrices up to an arbitrary complex, orthogonal matrix, whose elements are not even bounded. This situation contrasts with the mLRSM, since within this framework the imposition of a discrete LR symmetry is sufficient to fix  the arbitrary orthogonal matrix in terms of the heavy and light neutrino mass matrices (see for instance Refs.~\cite{Nemevsek:2012iq,Senjanovic:2016vxw,Senjanovic:2018xtu}).
%{\color{magenta} need statement here as to how the imposition of LR symmetry eliminates the arbitrariness of the $\mathcal{R}$ matrix}

The Dirac Yukawa coupling of heavy neutrinos could in principle be probed by considering angular asymmetries in the channel with one right-handed charged lepton, one left-handed charged lepton and two jets in the final state~\cite{Nemevsek:2012iq}.  In practice, as we shall show,  this may not be feasible with the statistics one expects even at the 100 TeV machine.   The reason is that for a large number of events $N$, the statistical error goes as  $\sqrt{N}$ and hence the asymmetry one wishes to measure should be at least of the order of  $\sqrt{N}/N = 1/\sqrt{N}$. {For instance, for the mLRSM and in the best case scenario    with $W_R$ boson mass of 6 TeV, the branching ratio of the HNs to charge leptons and one $W$ boson  is of the order of $10^{-4}$. Hence,  probing this small branching ratio would require at least $10^{8}$  signal events, which is not feasible at either the LHC  with high luminosity or a next 100 TeV $pp$-collider.  }   For a complete phenomenological study of the two leptons and two jets channel, see {\it e.g.} Refs.~\cite{PhysRevD.62.013001,Han:2012vk}.  

Consequently, in this work we propose instead that the ideal channel for probing the Dirac mass term of heavy neutrinos is through the purely leptonic decay  $W_R^{\pm} \rightarrow l^{\pm} N \rightarrow l^{\pm} (N \rightarrow l'^{\pm} W^{\mp}\rightarrow l^{\pm} l'^{\mp} \nu ) $. This channel  has been previously studied in the context of the type I  see-saw extension of the SM with fermion singlets in Ref. \cite{Izaguirre:2015pga}  and more recently searched for by the CMS collaboration \cite{Sirunyan:2018mtv}.  It has also been  studied  in the context of  a left-right symmetric model with an inverse see-saw mechanism in Ref. \cite{Das:2016akd} .

It provides a cleaner signal that has an advantage with respect to the KS channel since no asymmetry needs to be measured. Our main findings may be summarized as follows: in order to test the mLRSM prediction for the mixing between the heavy and light neutrinos (or equivalently the Dirac mass), one must consider a next generation collider beyond the LHC, such as a 100 TeV $pp$ collider. On the other hand, and if one observes evidence for this mixing at the LHC, it would point to a non-minimal scenario within the context of LRSMs.

%{\color{magenta} I think we need a brief summary of the main findings in the mLRSM and non-minimal LRSM here}

The discussion of our study leading to these findings is organized as follows. In Sec.~\ref{intro}, we briefly introduce the model in both its minimal and non-minimal incarnations.
%Then, in Sec.~\ref{HLsection}   
Assuming charge conjugation as the left-right symmetry for the minimal case,  we explain the relationship between the Dirac mass matrix in terms of the heavy and light neutrino mass matrices.  This connection is crucial to obtain the sensitivity to the Dirac mass.  In Sec.~\ref{sensitivities}, we estimate the sensitivity at the LHC, High Energy LHC (HE-LHC) and a 100 TeV $pp$ collider to heavy-light neutrino mixing. We compare the reaches of these various colliders to the expectations within the mLRSM and non-minimal model discussed in Section \ref{intro}. In Sec.~\ref{discussion} and within the minimal model, we translate the sensitivity to the heavy-light neutrino mixing into a reach on the Dirac mass. Finally in Sec.~\ref{conclusions} the conclusions are given.

\section{The left-right symmetric model}\label{intro}

%%%%%%%%%%%
\subsection{The minimal left-right symmetric model}
The minimal left-right symmetric model~\cite{Pati:1974yy,Mohapatra:1974gc,Senjanovic:1975rk} was introduced in order to explain the smallness of neutrino mass in connection with the  spontaneous violation of Parity~\cite{Senjanovic:1978ev,Mohapatra:1979ia,Mohapatra:1980yp}. In this work we do not pursue the ${\cal O}(1)$  Yukawa couplings of neutrinos, that is, without special Yukawa texture~\cite{Kersten:2007vk,deGouvea:2007hks,Xing:2009in,He:2009ua,Ibarra:2010xw,Haba:2011pe,Mitra:2011qr,Chen:2013fna} or cancellation between Type I and Type II see-saw effect~\cite{Akhmedov:2006de,Akhmedov:2006yp,Chao:2007mz}.
In this case, the LR symmetry breaking scale would be very high, such that the $W_R$ boson and heavy neutrino will be too heavy to be produced even in a future 100 TeV collider. Instead,  we consider the minimal framework, where we have relatively small Yukawa couplings (${\cal O}(10^{-6}\sim10^{-5})$) with relatively light $W_R$ boson mass (${\cal O}(1\sim10)$TeV) and in the reach of present and future colliders.

\textbf{The gauge group and field content:} the gauge group is  $\mathcal{G}=SU(2)_L\times SU(2)_R\times U(1)_{B-L}$, with an additional discrete symmetry that may be generalized parity ($\mathcal{P}$) or  charge conjugation ($\mathcal{C}$).  The quarks and leptons are  doublets in the following irreducible representations of the gauge group:
 \begin{align*}
q_L =
\left( \begin{array}{ccc}
u  \\
d \\
\end{array} \right) _L :   (2,1,\frac{1}{3}), \quad  q_R
=\left( \begin{array}{ccc}
u  \\
d \\
\end{array} \right) _R :   (1,2,\frac{1}{3}),
\end{align*}

\begin{align}
\
L_L =
\left( \begin{array}{ccc}
\nu  \\
l \\
\end{array} \right) _L :   (2,1,-1),\quad  L_R =\left(
\begin{array}{ccc}
N  \\
l \\
\end{array} \right) _R :  (1,2,-1). \nonumber \\
\end{align}
Where $N$ represents the new heavy neutrino states, whose presence explain the smallness  of neutrino masses on the basis of the see-saw mechanism~\cite{Minkowski:1977sc,Mohapatra:1979ia,Glashow:1979nm,GellMann:1980vs,Yanagida:1979as,Schechter:1980gr}.  

\textbf{The Higgs sector sector of the mLRSM} \cite{Minkowski:1977sc,Mohapatra:1979ia}, consists in   one bidoublet $\Phi$, in the (2,2,0) representation of $\mathcal{G}$ and  two scalar triplets $\Delta_L$ and $\Delta_R$, belonging to (3,1,2) and (1,3,2) representation respectively
 \begin{align}
\Phi =
\left( \begin{array}{ccc}
\phi_1^0 && \phi_2^+  \\
\phi_1^- && \phi_2^{0 }\\
\end{array} \right) , \quad  \Delta_{L,R}=\left( \begin{array}{ccc}
\delta^+_{L,R}/\sqrt{2} && \delta_{L,R}^{++}  \\
\delta_{L,R}^{0} &&- \delta^+_{L,R}/\sqrt{2}  \\
\end{array} \right). \nonumber \\
\end{align}

After SSB,  the v.e.v's of the Higgs fields may be written as \cite{Mohapatra:1980yp}
 \begin{eqnarray}
\langle \Phi \rangle =
\left( \begin{array}{ccc}
v_1 && 0  \\
0 && v_2 e^{i\alpha} \\
\end{array} \right),
\end{eqnarray}

 \begin{eqnarray}
\langle \Delta_{R} \rangle=\left( \begin{array}{ccc}
0&&0 \\
v_R  &&0  \\
\end{array} \right)  , \quad  \langle \Delta_{L} \rangle=\left( \begin{array}{ccc}
0&&0 \\
v_Le^{i\theta_L} &&0  \\
\end{array} \right),
\end{eqnarray}
 where  $\alpha$ and $\theta_L$ are called the ``spontaneous'' CP  phase and  {$v_L \ll   v_1^2+v_2^2 \ll   v_R^2$}. All the physical effects due to $\theta_L$  can be neglected, since this phase is always accompanied by the small $v_L$.

 Under the discrete left-right symmetry the fields transform as follows:
\begin{align}
\mathcal{P} :  \left\{ \begin{array}{ll}
 \mathcal{P}f_L\mathcal{P}^{-1} =  \gamma_0f_R  \\
 \mathcal{P}\Phi\mathcal{P}^{-1} =\Phi^{\dagger} \\
 \mathcal{P}\Delta_{(L,R)}\mathcal{P}^{-1} = -\Delta_{(R,L)}  %\\
% \mathcal{P}\Delta_R\mathcal{P}^{-1} = -\Delta_L
\qquad
 \end{array} \right. \quad \mathcal{C} :  \left\{ \begin{array}{ll}
 \mathcal{C}f_L  \mathcal{C}^{-1} =C(\bar{f_R})^T  \\
 \mathcal{C}\Phi\mathcal{C}^{-1} = \Phi^{T} \\
 \mathcal{C}\Delta_{(L,R)}\mathcal{C}^{-1} = -\Delta^*_{(R,L)} %\\
 % \mathcal{C}\Delta_R\mathcal{C}^{-1} = -\Delta^*_L
 \end{array} \right.  \label{relations}
 \end{align}
 where $\gamma_{\mu}$ ($\mu=0,1,2,3.$) are the gamma matrices and $\mathcal{C}$ is the charge conjugation operator.

 \textbf{Lepton masses:} lepton masses are due to the following Yukawa interactions (once the Higgs fields take their v.e.v along their neutral components)
\begin{eqnarray}
\mathcal{L}_Y=&\bar{L}_L(Y_{\Phi}\Phi+\tilde{Y}_{\Phi}\tilde{\Phi})L_R +\frac{1}{2}(L_L^TCi\sigma_2Y_{\Delta_L}\Delta_LL_L \nonumber \\ 
&+L_R^TCi\sigma_2Y_{\Delta_R}\Delta_RL_R )+h.c., \label{yukawas}
\end{eqnarray}
where $\tilde{\Phi}=\sigma_2\Phi^*\sigma_2$ , $\sigma_2$ is the Pauli matrix and $C\equiv i\gamma_2\gamma_0$.

 Invariance of the Lagrangian under the Left-Right symmetry requires the Yukawa couplings to satisfy
 \begin{eqnarray}
 \mathcal{P} :  \left\{ \begin{array}{ll}
 Y_{\Delta_{R,L} } = Y_{\Delta_{L,R}} \\
 Y_{\Phi}=Y_{\Phi}^{\dagger} \\
 \tilde{Y}_{\Phi}=\tilde{Y}_{\Phi}^{\dagger}
 \end{array} \right.,\quad
\mathcal{C} :  \left\{ \begin{array}{ll}
 Y_{\Delta_{R,L} } = Y^*_{\Delta_{L,R}} \\
 Y_{\Phi}=Y_{\Phi}^T \\
 \tilde{Y}_{\Phi}=\tilde{Y}_{\Phi}^T \label{LRcond}
 \end{array} \right.
 \end{eqnarray}
 
 Consistent with the above notation, the neutrino mass matrix of neutrinos is of the form \cite{Mohapatra:1979ia,Mohapatra:1980yp}
 \begin{equation}
 \mathcal{L}_{\nu} =\frac{1}{2}\left( \begin{array}{ccc}
\nu  & N^c \\
\end{array} \right)_L^T C
 \left( \begin{array}{ccc}
 M_L &M^{*}_D  \\
 M_D^{\dagger} & M_{R} \\
\end{array} \right)
\left( \begin{array}{ccc}
\nu  \\
N^c \\
\end{array} \right) _L+ h.c.,
 \end{equation}
 where $N_L^c \equiv C \bar{N}_R^T$ and $M_L$, $M_R$ and $M_D$ are $3\times 3$ matrices given by
\begin{eqnarray}
& M_{L}\equiv Y_{\Delta_L}v_Le^{i\theta_L}, \\
&M_R \equiv Y_{\Delta_R}^*v_R,\\
&M_D\equiv v_1Y_{\Phi}+\tilde{Y}_{\Phi}v_2 e^{-i\alpha}.
\label{masses}
\end{eqnarray}
After  diagonalization, the light and heavy neutrino mass matrices takes the see-saw form:
\begin{eqnarray}\label{numasses}
&M_{\nu}\simeq M_L-M^{*}_D\frac{1}{M_N}M_D^{\dagger}, \label{numass} \\
& M_{N}\simeq M_R.
\end{eqnarray}
The contributions to the light neutrino masses proportional to  $M_D$  and $M_L$ are called the Type I and Type II see-saw contributions respectively. It follows from the seesaw formula that the    eigenstates corresponding to Eqs.~\eqref{numasses} are given by
 \begin{align}
\left( \begin{array}{ccc}
\nu'  \\
N'^c \\
\end{array} \right) =
\left( \begin{array}{ccc}
 1 &\Theta   \\
-\Theta^T & 1 \\
\end{array} \right)  
\left( \begin{array}{ccc}
\nu  \\
N^c \\
\end{array} \right) ,
\end{align}
where  the heavy-light neutrino mixing is given by
 \begin{equation}
\Theta \simeq  M_D^{*}M_N^{-1}. \label{HLmixing}
 \end{equation}
Finally, the charged lepton mass matrix is given by
\begin{eqnarray}
&M_l= Y_{\Phi}v_2e^{i\alpha}+\tilde{Y}_{\Phi}v_1.
\end{eqnarray}
As usual, the mass matrices can be diagonalized  by the bi-unitary transformations
\begin{eqnarray}
&M_l = U_{lL}\hat{M}_lU^{\dagger}_{lR}, \nonumber \\ 
 &M_{\nu}=U^{*}_{\nu }\hat{M}_{\nu}U^{\dagger}_{\nu},\quad M_{N}=U^{*}_{N }\hat{M}_{N}U^{\dagger}_{N},
\end{eqnarray}
where $\hat{M}_l$, $\hat{M}_{\nu}$ and $\hat{M}_{N}$ are diagonal matrices with real, positive eigenvalues.

\textbf{Charged gauge  interactions with leptons:} from the covariant derivative and  in the mass eigenstates  basis,  the charged current Lagrangian is
\begin{equation}
\mathcal{L}_{cc}= \frac{g}{\sqrt{2}}(\bar{l}_L V_{L} \slashed{W}_{\!L} \nu_L -\bar{l}_L \Theta_L \slashed{W}_{\!L} N_L^c +
\bar{l}_R V_{R} \slashed{W}_{\!R} N_R+ \bar{l}_R \Theta_R \slashed{W}_{\!R} \nu^c_R)  +h.c.,\label{cclagrangian}â
\end{equation}
where  $N_R \equiv C (\bar{N^c_L})^T= i\gamma_2\gamma_0(N_L^{c})^*$, $\nu_R^c \equiv C(\bar{\nu}_L)^T$ and  $\gamma_0$ and $\gamma_2$ are the gamma matrices and the mixing matrices  $V_L$, $V_R$, $\Theta_L$ and $\Theta_R$ are  given by
\begin{eqnarray}                                                                 
& V_L = U^{\dagger}_{lL}U_{\nu},\quad
\Theta_L= U_{lL}^{\dag}\Theta U_{N} 
\\
& V_R = U^{\dagger}_{lR}U^*_{N},
\quad
 \Theta_R =  U_{lR}^{\dag} \Theta^{\dag} U_{\nu}^*
\end{eqnarray}

We may use the freedom of rephasing the charged lepton fields to remove three unphysical phases from $V_L$, which ends up having 3 mixing angles and 3 phases,  {namely one Dirac and two Majorana phases}.  {Since the freedom of rephasing the charged lepton is already used for $V_L$, its right-handed analog --the leptonic mixing matrix $V_R$-- is a general $3\times 3$ unitary matrix and may be therefore parametrized by 3 mixing angles and 6 phases.} 

A comment regarding the mixing matrices $\Theta_L$ and $\Theta_R$ is in order: for charge conjugation as the LR symmetry, without loss of generality  one can choose $U_{lL}=U_{lR}=1$, such that $V_L=U_{\nu}$ and $V_R=U_N^*$. In this case, the mixing matrices can be written in the form   
\begin{align}\label{theta_phys_P}
\Theta_L= \Theta V_R^*,\quad \Theta_R= \Theta V_L^*.
\end{align}

For Parity as the LR symmetry, it is no longer true  that one can assume $U_{lL}=U_{lR}=1$. Nevertheless, since the Dirac mass matrix is  hermitian with  a very good approximation,  even in this case one can write 
\begin{equation}\label{theta_phys_C}
 \Theta_L= \Theta V_R^*\left[1+\mathcal{O}\left(\hat{M}_l\tan2\beta \sin \alpha\right)\right],\quad   \Theta_R= \Theta V_L^* \left[1+\mathcal{O}\left(\hat{M}_l\tan2\beta \sin \alpha\right)\right]. 
\end{equation}
 where the parameter $\tan2\beta\sin\alpha\lesssim 2m_b/m_t$~\cite{Senjanovic:2014pva,Senjanovic:2015yea},  $\beta\equiv v_2/v_1$,   $m_b$ and $m_t$ are the bottom and top quark masses respectively. Therefore, up to small terms of the order $\mathcal{O}\left(\hat{M}_l\tan2\beta \sin \alpha\right)$, the heavy light mixing matrices are roughly  the same for both   parity and charge conjugation as the LR symmetry. 

%%%%
\textbf{Heavy-light mixing in the mLRSM: }\label{HLsecton}
in the mLRSM, the heavy-light neutrino mixing depends on the light and the heavy neutrino {mass matrices}. It is known~\cite{Nemevsek:2012iq} that this mixing enters in the decay of the heavy neutrino into a left-handed charged lepton and two jets~\cite{Keung:1983uu}, and one could measure the mixing by measuring the chirality~\cite{Han:2012vk} of the outgoing charged lepton in order to discriminate this channel from the usual channel where the heavy neutrino decays into a right-handed charged leptons and two jets\footnote{The relative strength of these two channels can be seen for instance from the "phase diagram" of the heavy-light mixing in the see-saw models in Ref.~\cite{Chen:2013fna}, and it is pointed out that without special Yukawa texture the ordinary channel with heavy neutrino decay with right-handed charged current is generally much larger than the left-handed current.}.  Instead, in this work we point out that the same mixing enters in the purely leptonic decay of the heavy neutrinos. This channel has an advantage with respect to the channel with two leptons and two jets, since no asymmetry \textcolor{black}{(chirality information)} needs to be measured in order to obtain the heavy-light mixing matrix elements. %\textcolor{blue}{Notice that  the  heavy neutrino decay $N\to \mu^- W_R^{(*)}\to \mu^- e^+ \nu_e$   is suppressed by an additional power of  the heavy-light mixing and  therefore   it can    be  neglected at the leading order.}  \textcolor{red}{ JCV. This phrase was somewhat  misleading since $W_R$ doesn't decay into light neutrinos  at the leading orde\color{magenta} the foregoing was somewhat unclear}. 
In addition, from the experimental perspective, the backgrounds relevant to the purely leptonic channel are cleaner.  This channel has been previously studied in  Refs.~\cite{Izaguirre:2015pga,Dube:2017jgo} including  both prompt and displaced vertex for the signal in the context of the SM extended by a fermion singlet. \textcolor{black}{In the latter instance, no heavy resonance $W_R$ is produced in the process, which makes kinematics for the final states very different with respect to the present work}. 
%In principle one could measure the chirality~\cite{Han:2012vk} of the outgoing charged lepton in order to discriminate this channel from the usual channel with two charged leptons and two jets. Instead,  we propose to study the purely leptonic channel, whose amplitude is proportional to the heavy-light mixing in Eq.~\ref{HLmixing}. From the experimental perspective, this channel is cleaner than the hadronic decay. 

 In what follows and  for the sake of illustration, we consider $\mathcal{C}$ as the LR symmetry but the same conclusions hold for the case when the LR symmetry corresponds to $\mathcal{P}$. From Eq. \eqref{LRcond} it follows that the Dirac mass term is symmetric  and Eq.~\eqref{numass} takes the form~\cite{Nemevsek:2012iq}
\begin{equation}
M_{\nu}\simeq Y_{\Delta_L}v_Le^{i\theta_L}-M^{*}_D\frac{1}{M_N}M_D^{*}.
\end{equation}
Multipling from the left by $M_N^{-1}$ one gets~\cite{Tello:2012qda}
\begin{align}
M_N^{-1}M_{\nu} & \simeq M_N^{-1}Y_{\Delta_L}v_Le^{i\theta_L}-\frac{1}{M_N}M^{*}_D\frac{1}{M_N}M_D^{*}, \\
M_N^{-1}M_{\nu} & \simeq M_N^{-1}M_L-\Theta^2.
\end{align}
Hence, the mixing angle can be written in terms of the heavy and light neutrino masses as~\cite{Tello:2012qda} 
\begin{equation}
\Theta = \sqrt{\epsilon-M_N^{-1}M_{\nu} }= M_D^{*}M_N^{-1},
\label{ThetaHL_phys}
\end{equation}
with $\epsilon\equiv v_L/v_R$.  See Ref. \cite{Senjanovic:2016vxw,Senjanovic:2018xtu} for the determination of the analogue of Eq.~\eqref{ThetaHL_phys}  for Parity as the LR symmetry.  In what follows and for the sake of simplifying the discussion,  we set $v_L=0$, effectively assuming   type I see-saw dominance for the light neutrino masses. Notice that the choice $v_L$ small  is technically natural,  as discussed originally in Ref.~\cite{Mohapatra:1980yp} and more recently revisited in Ref.~\cite{Maiezza:2016ybz}. Finally,  notice that the mixing matrix $\Theta$  (equivalently $M_D$) is  a complex matrix, so that no issues arrise due to the $-1$ factor inside the square root and the fact that $\epsilon$ is a complex quantity. In any case,  this phase  phase factor has no impact in our analysis below. For a discussion of its physical significance see, for instance, Ref.~\cite{Tello:2012qda}. 

In the next section and using the above leptonic channel,  we study the sensitivity of the LHC, HE-LHC and a 100 TeV $pp$ collider, to the mixing in Eq. \eqref{ThetaHL_phys} as a function of  $M_{W_R}$ and the { lightest} heavy neutrino mass $m_{N}$, for benchmark values of the other heavy neutrinos.  Later, from this sensitivity and using Eq. \eqref{HLmixing}, one can infer the values of $M_D$ that can be probed at the LHC and the next generation of hadronic colliders.
 \begin{figure}\centering
     \includegraphics[scale=0.3]{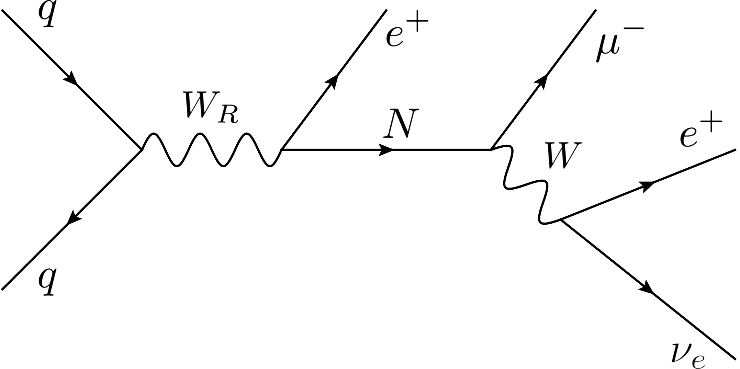} 
     \quad
      \includegraphics[scale=0.3]{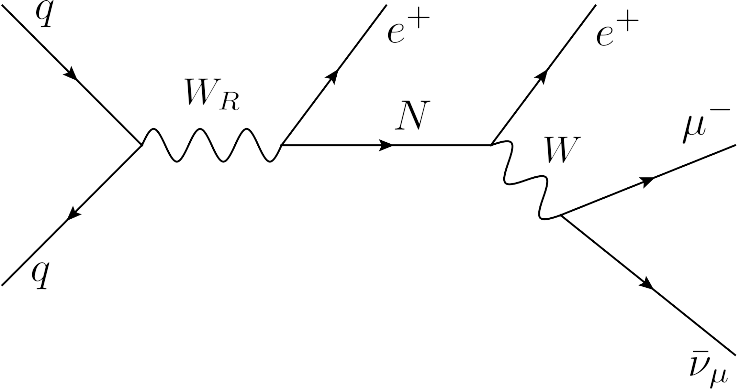}
  \caption{ Feynman diagrams for the $W_R$ production and  leptonic decay $N\rightarrow e^+ \mu^- \nu$, where $\nu$ can be either a $\nu_e$ or $\bar{\nu}_{\mu}$.  }
  \label{fig:KSlep}
\end{figure}

\subsection{``Non minimal" left-right symmetric model}
%In this section we will discuss the possibility to have a non-minimal LR scenario in which the neutrino masses are generated by the inverse seesaw mechanism.

As explained in the previous section,  in the mLRSM  the gauge group is broken to the SM group through the triplet  $\Delta_R$ and the bidoublet scalar $\Phi$. This construction generates a seesaw mass for the right-handed neutrinos from the vacuum expectation value of the $\Delta_R$. Here, we will consider a slightly different  LR scenario, a "non-minimal" model,  in which now the LR group is broken through  a  doublet scalar { in the (1,2,-1) representation of $\mathcal{G}$ } ~\cite{Parida:2010wq, Arbelaez:2013nga}. Adding an extra vector of gauge singlet fermions  $S = (S_1,S_2,S_3)^T$ to the particle content the neutrino masses will be generated now by an inverse seesaw mechanism~\cite{Mohapatra:1986bd, Dev:2015pga}. The inverse seesaw scenario in the context of left-right symmetry was studied in detail in Ref. \cite{Anamiati:2016uxp}. In this section, we will only review the most important results. 
%{\color{magenta} the notation seems a bit counterintuitive to me: $\Psi$ are scalars and $S$ are fermions. Is this the standard convention for this model?}.

We work in the basis in which the charged lepton mass matrix  is diagonal. The inverse seesaw neutrino mass matrix in the interaction basis for the neutral states ${\cal N} = (\nu_L, N^c,S^c)^T$ can be written in a  $3 \times 3$ notation as:
\begin{align}
{\cal M} =
 \label{Mnu}
\left( \begin{array}{ccc}
 0 &M^{T}_D& 0  \\
 M_D & 0 & M_N \\
 0 &  M^{T}_N & \mu 
\end{array} \right)\ \,
\end{align}
where $M_D$, $M_N$, and $\mu$ denote $3 \times 3$ matrices and 
the sub-matrix $\mu$ is taken to  be diagonal.
%\footnote{The sub-matrix $\mu$ can always be brought to a diagonal from  after a redefinition of the gauge singlets S}.
% and  the Majorana sub-matrix $\mu$ can always be brought to a diagonal from \cite{Anamiati:2016uxp}.
 Assuming the sub-matrices $M_D$, $M_N$, $\mu$  have mass scales arranged hierarchically, $M_N \gg M_D, \mu$, the light neutrino mass matrix $M_{\nu}$  can be expressed in terms of the  matrices in Eq. (\ref{Mnu}) as
 %{\color{magenta} notation is a bit confusing. Above, $M_N$, $M_D$ and $\mu$ are matrices. In the last expression there is some mixed notation}
 %, integrating out the heavy RH-neutrinos leads to the %effective neutrino mass matrix.
%
\begin{eqnarray}
M_{\nu}\simeq M^{T}_D\frac{1}{M^T_N} \mu \frac{1}{M_N}  M_D.
 \label{numass2} 
\end{eqnarray}
%
%Being symmetric by construction, $M_{\nu}$ can be diagonalized as
%\begin{eqnarray}
%\hat{M}_\nu}= V^{T}_L m_\nu  V_L
% \label{numass2} 
%\end{eqnarray}
  %
 %This equation shows that in an inverse seesaw scenario the %suppression of the light neutrino masses can be attributed to  %small values of $\mu$ without the need of exceedingly small %values of $M_D$.  
  
  Using  the bi-unitary transformations 
\begin{eqnarray}
&M_{\nu}=V^{*}_{L }m_{\nu}V^{\dagger}_{L},\quad M_{N}=V_{R }\hat{M}_{N}U^{\dagger}_{R},
 \label{numass0}
\end{eqnarray}
the mass matrix ${\cal M }$ can be diagonalized into 
\begin{align}
{\cal \hat{M} } =
\label{numassh}
\left( \begin{array}{ccc}
 m_\nu &0& 0  \\
 0 & \hat{M}^{-}_N & 0 \\
 0 &  0 & \hat{M}^{+}_N 
\end{array} \right).
\end{align}

%of the mass eigenstates  $N = (\nu, N_{-}, N_{+})^T$

   Here $\hat{M}^{-}_N, \hat{M}^{+}_N $ and $\hat{M}_N $ are  diagonal mass matrices with   $ \hat{M}^{\pm}_N = \hat{M}_N  \pm \frac{1}{2}\mu^V$ and $\mu^V = V^T_R \mu V^T_R $. The neutral mass eigenstates ${\cal N^\prime}= (\nu, N_{-}, N_{+})^T$ correspond to  three light neutrinos and  three    pairs of almost  degenerate heavy neutrinos   with mass eigenvalues   $m_{N^\pm_{i}}= (\hat{M}^\pm_N)_{ii}= (\hat{M}_N)_{ii} \pm \frac{1}{2}(\mu^V)_{ii}$. %The heavy mass eigenvalues form  three    %pairs of almost  degenerate neutrinos, each %pair  with masses $m_{N_i}=(\hat{M}_N)_{ii} $ %split by the small quantity  $(\Delta M)_{i} %= (\mu^V)_{ii} $ (for $i = 1,2,3$).
   %{\color{magenta} again unclear notation; also is one mass $M_N$ and the other two $M_N\pm\Delta M$ -- it is not clear}
   \footnote{Here the three pairs of  -- almost degenerate -- neutrinos correspond to the so-called "quasi-Dirac" neutrinos \cite{Anamiati:2017rxw,Anamiati:2016uxp, Das:2016akd}.
}
%\cite{Anamiati:2016uxp}.

 Using Eqs (\ref{numass2}),  (\ref{numass0}) the light neutrino mass matrix can be written as
\begin{eqnarray}
m_{\nu}= V^T_{L} M^{T}_D\frac{1}{M^T_N} \mu \frac{1}{M_N}  M_D V_{L}.
 \label{numass3} 
\end{eqnarray}
Following the parameterization developed by Casas and Ibarra \cite{Casas:2001sr} we can now write $M_D$ as: 
\begin{eqnarray}
M_D = M_N \frac{1}{\sqrt{\hat{\mu}}} {\cal R} \sqrt{m_\nu} V^\dagger_{L}.
\end{eqnarray}
 Here the matrix ${\cal R}$ is an arbitrary  complex orthogonal matrix.    Rewriting $M_N$ using Eq. (\ref{numass2}) one finds:
 \begin{eqnarray}
V_R^\dagger M_D = \hat{M}_N U_R^\dagger \frac{1}{\sqrt{\hat{\mu}}} {\cal R} \sqrt{m_\nu} V^\dagger_{L},
\label{sesaw}
\end{eqnarray}
which express $V_R^\dagger M_D$ in terms of the low energy observables $m_\nu$, $V_L$ allowing us to reproduce the neutrino data.   {  Notice that in practice,  the arbitrariness of the  matrix $\cal{R}$  is a consequence of the fact that for the non-minimal models,  the Dirac mass matrix is arbitrary.} { This feature precludes a direct mapping of neutrino data onto $M_D$ in non-minimal models.}

 %{\color{magenta} should emphasize the presence of $\mathcal{R}$ prevents a direct mapping of the neutrino data onto $M_D$}
 %For simplicity, we have set in our numerical calculations the arbitrary matrices ${\cal R}, U_R$ to the identity ${\cal R} = U_R =I$.   

 The mixing matrix ${\cal V}$ that relates  the neutral mass eigenstates ${\cal N^\prime}$and the interaction eigenstates ${\cal N}$ via ${\cal N} = {\cal V} \ {\cal N^{\prime}}$      can be expressed in the seesaw approximation as \cite{Anamiati:2016uxp}:
\begin{align}
{\cal V} \simeq
\left( \begin{array}{ccc}
 V_L & i\Theta_L& \Theta_L  \\
 0 & -\frac{i}{2} V^*_R & \frac{1}{2} V^*_R \\
 -\sqrt{2} U_R \Theta_L^{\dag} V_L & \frac{i}{\sqrt{2}} U_R& \frac{1}{\sqrt{2}} U_R
\end{array} \right),
\label{mixing}
\end{align}
 where   $\Theta_L = \frac{1}{\sqrt{2}} M_D^{\dagger}  V_R \hat{M}_N^{-1}$. 
 %is  proportional to $M_D$, and in turns,   all the couplings of %heavy neutrinos with the SM  gauge bosons are also proportional $M_D$ 
 \footnote{The expressions of the couplings of the heavy neutrinos to the gauge bosons are given in  \cite{Anamiati:2016uxp}.}
%Since  in an inverse seesaw scenario $M_D$ its not required to %be exceedingly small, large branching ratios $Br(N \rightarrow % l W)$ are expected for  small values of $\mu$. 

%The charge current lagrangian is
%
%\begin{equation}
%\mathcal{L}_{cc}= \frac{g}{\sqrt{2}}(\bar{\nu}_L V_{L}^\dag \slashed{W}_{\!L} l_L+(\bar{N}_+ i \bar{N}_-) V_{R}^\dag \slashed{W}_{\!R} l_R) +h.c.,\label{cclagrangian2}â
%\end{equation}
% $V_L$ and $V_R$ are the left and right leptonic mixing matrices respectively
%\begin{eqnarray}                                                                 
%& V_L = U^{\dagger}_{lL}U_{\nu}, \\
%& V_R = U^{\dagger}_{lR}U_{N}.
%\end{eqnarray}

\section{Collider sensitivities} \label{sensitivities}

As discussed in the previous section, the most promising channel for the determination of the Dirac Yukawa coupling of neutrinos is the purely leptonic channel $p p \rightarrow W_R^{\pm} \rightarrow l^{\pm} l^{\pm} l^{\mp} \nu $.
For purposes of illustration, we focus on the process  $p p \rightarrow e^+ N\rightarrow e^+  \mu^- e^+ \nu $ (see Fig.~\ref{fig:KSlep}) rather than 
$p p \rightarrow \mu^+ N\rightarrow \mu^+  e^- e^+ \nu $ in order to avoid the presence of an $e^+e^-$ pair in the final state.
The final state with different flavors for leptons of the same charge has a cleaner Standard Model background and also avoids events coming from the heavy neutrino decaying through the neutral currents (for example  $p p\to W_R^+\to e^+ N\to e^+ \nu Z^*_{(R)}\to e^+ e^+e^-\nu $).

We study the main sources of background for the process $ p p \rightarrow e^+ N\rightarrow e^+  \mu^- e^+ \nu $ for different center of  mass energies. 
%This particular final state has only been chosen for the sake of illustration.  
In what follows,  we discuss  the LHC expected sensitivity to the branching ratio of HNs decaying  into leptons at the LHC  with $\sqrt{s}= 13$ TeV, the high energy LHC (HE-LHC) with $\sqrt{s}= 28$ TeV  and a $pp$ collider with $\sqrt{s}= 100$ TeV. We compare  our cross section results with those obtained in Refs.~\cite{Nemevsek:2012iq,Senjanovic:2016vxw,Mitra:2016kov,Nemevsek:2018bbt} for the $p p \rightarrow e^+ N$ production and find agreement. 

 Assuming that the neutrinos in the final state cannot be distinguished, the decay width  of heavy neutrinos into three leptons $\Gamma(N\rightarrow l^{\pm}l'^{\mp}\nu)$ is proportional to the heavy-light mixing and it is of the form  
\begin{equation} \label{leptonicwidth}
\Gamma\left(N\rightarrow l^{\pm}l'^{\mp}\nu\right) = \left(|(\Theta_L)_{lN}|^2+|(\Theta_L)_{l'N}|^2\right) \frac{G_F^2}{96\pi^4m_N}\int_0^{m_N^2} dx  \frac{\pi(m_N^2-x)(m_N^4+xm_N^2-2x^2)}{m_N^2(1-\frac{x}{M_W^2})^2}.
\end{equation} 

 \begin{figure}\centering
     \includegraphics[scale=0.4]{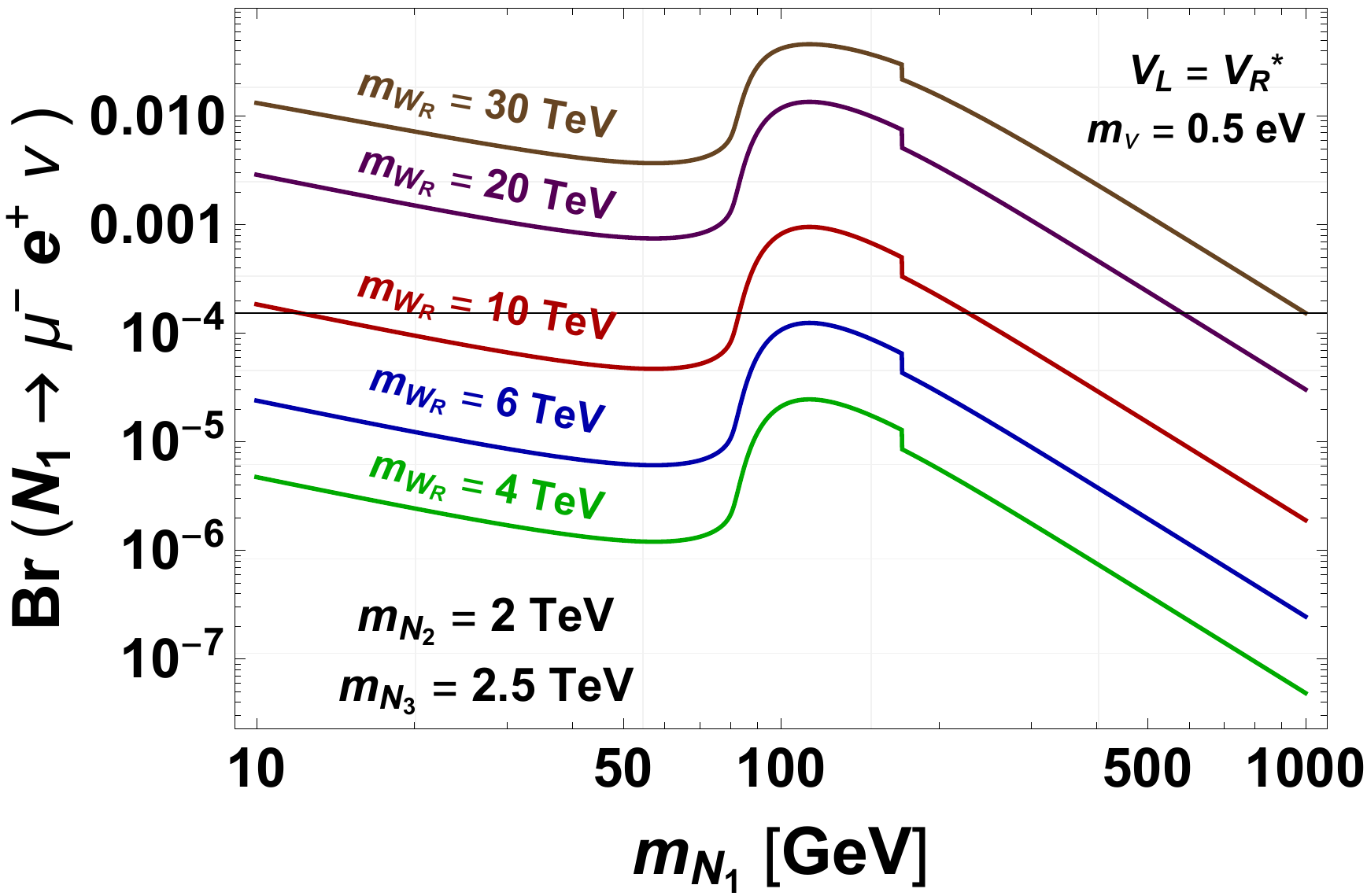}
  \caption{Branching ratio of the purely leptonic decays of the heavy neutrino $N$ in the minimal left-right symmetric model.  We use the indicative limit for light neutrino masses of $m_{\nu}=0.5$ eV~\cite{Nakamura:2010zzi}.}
  \label{fig:br2lep}
\end{figure}
{Where $m_N$ denotes the mass of the heavy neutrino.\footnote{{{In the inverse seesaw scenario,  $m_N$ denotes collectively the pair of mass eigenvalues $m_{N^\pm}$ for $N = N_{\pm}$}}.} For illustration we assume $V_L=V_R^*$ and the indicative upper limit on light neutrino masses  $\sum_{\nu} m_{\nu}=0.5$ eV~\cite{Nakamura:2010zzi}.  In Figure \ref{fig:br2lep} we show the branching  ratio of  the heavy neutrino $N$ into $e^+\mu^-\nu$ as a function of the lightest heavy neutrino (HN) mass {{ in  the minimal left-right symmetric model}.
   As can be seen from the figure, the branching ratio into leptons decreases as the heavy neutrino mass $m_N$   increases.  This feature is due to the proportionality of the leptonic branching ratio to $\Theta_L^2$ (see Eq.~\eqref{leptonicwidth}), which in turn is proportional to $1/m_N$ --see Eq.~\eqref{ThetaHL_phys}.
   
   Another important feature is the increase of the leptonic branching ratio as the $W_R$ boson mass increases. This occurs because the dominant process with one lepton and two jets has an additional suppression of $M_{W_R}$. The net effect is to make the branching ratio into leptons increase when the $W_R$ boson mass increases.  Finally, the bump when $m_N\sim    M_W$ is due to the transition from three body decay to a two-body decay through an on-shell $W$ boson. The drop in the decay rate due to the top quark threshold is also evident.

{Regarding the  processes shown in Fig.~\ref{fig:KSlep},  we find  two   issues that  may affect the selection  efficiency of the signal:  (1) the two origins of the $\mu^-$, which is an interpretation issue and (2) the possible jet fake background}: 
\begin{enumerate}
\item  The origin of the $\mu^-$: there are two possibilities
\begin{equation}
  p p \rightarrow e^+ N \rightarrow  e^+  \mu^- (W^+\rightarrow e^+ \nu_e ), \label{posi1}
\end{equation}
and
\begin{equation}
p p \rightarrow e^+ N \rightarrow e^+   e^+ (W^- \rightarrow \mu^- \bar{\nu}_{\mu} ). \label{posi2}
\end{equation}
Namely, the final state muon can be directly produced in the decay of the heavy neutrino $N$ or it can also be produced in the decay of the $W$ boson that comes from the decay of the heavy neutrino $N$. Notice that the lepton flavor cannot be used to discriminate among the two processes since the light neutrino goes undetected.  

For $m_N > m_W$, the transverse mass of the subleading positron and missing transverse energy $\slashed{E}_T$ system $m_T(e^+_{sub}\slashed{E}_T)$ may be helpful for discriminating between the two processes. In the process in Eq.~(\ref{posi1})  the subleading positron comes primarily from the decay of an on-shell $W$ boson, so the transverse mass of the subleading positron and missing transverse energy system will have a sharp decline around the mass of the $W$ boson. On the contrary, in the process in Eq.~(\ref{posi2}), the subleading positron directly comes from the decay of heavy neutrino $N$, so one may expect a broader distribution of this transverse mass. In principle, then, implementing a cut on this transverse mass near $m_W$ should remove a significant portion of events from the process in Eq.~(\ref{posi2}) while retaining most events coming from the process in Eq.~(\ref{posi1}).  We expect this method will be useful for $m_N \gg m_W$, but will be of less utility for $m_N$ near $m_W$. In the latter regime, the transverse mass mentioned above will be narrow in both processes, and, therefore, not efficient in discriminating between these two processes. We will return to this point below and show the distribution of this transverse mass for different benchmarks in Fig.~\ref{fig:MTe2met}.
%the $p_T$ distribution of the muon would be peaked at different values, and this might be used in principle to discriminate between the two channels. In practice and for the mass range of interest,  we find that this difference is smeared out after detector effects are included. 

For $m_N < m_W$  the decay goes through an off-shell $W$ boson. In this case, it seems at first glance more difficult to distinguish between these two contributions. In principle, one cannot determine the origin of the muon for a single event. However, a way to determine the proportion of $\mu^-$ from each channel by measuring a particular forward-backward asymmetry with an ensemble of events has been proposed in Ref.~\cite{Arbelaez:2017zqq}.
%in principle used in order to distinguish the origin of the muon. 
 %{\color{magenta} the following sentence seems completely disconnected from the discussion of the angular distributions. Why do we give a detailed explanation of this angular distribution method if we are not going to use it? (JCV) I agree with your comment I removed these expressions.} 
{In this work, we follow a different approach. In both cases mentioned above, when deriving the sensitivity to the decay branching ratio ${\rm Br}(N\to e^+ e^+ \mu^- \nu)$, we first estimate the efficiency  for each channel shown in Fig.~\ref{fig:KSlep} (see Fig.~\ref{fig:sigeff}) and subsequently compute  the  average  efficiency  by weighting each channel with the corresponding  probability of occurrence.

% We  first estimate the efficiencies  for each channel shown in Fig.~\ref{fig:KSlep} (see Fig.~\ref{fig:sigeff}) and subsequently compute  the  average  efficiency  by weighing each channel with their corresponding  probability of occurrence.
}

%\textcolor{blue}{In both cases mentioned above, When deriving the sensitivity on the decay branching ratio ${\rm Br}(N\to e^+ e^+ \mu^- \nu)$, We first estimate the efficiencies  for each channel shown in Fig.~\ref{fig:KSlep} (see Fig.~\ref{fig:sigeff}) and subsequently compute  the  average  efficiency  by weighing each channel with their corresponding  probability of occurrence. }

\item {Since the mixing parameter $\Theta_L$ may be quite small, there may be} a  non-negligible jet-fake background coming from the process $p p \rightarrow W_R^{\pm} \rightarrow l^{\pm} l^{\pm} j j  $, where one of the jets is misidentified as either an electron or a muon, since we did not reject extra jets in our analysis. Notice that the branching ratio for this channel is by far the dominant one. Therefore,  it can mimic the purely leptonic signal with one of the jets faking to leptons, which may be a contamination and decrease the sensitivity to the heavy-light mixing. {As discussed below, we address this issue}  by implementing cuts on the missing transverse energy $\slashed{E_T}$ and the transverse mass of the subleading electron and $\slashed{E_T}$ system. We find that for heavy neutrino masses $m_N < 1$ TeV, this channel is subdominant with respect to the trilepton channel. 
%{\color{magenta} But in the previous sentence, you mention that the additional hard jet could be a powerful discriminating tool. These cuts do not have anything obvious to do with the additional hard jet. Why are we not imposing a jet veto, then? There seems to be a logic \lq\lq disconnect" here. (JCV) I agree this is an old sentence, I erased that part of the jets. }
%due to the additional  $m_N/M_{W_R}$ suppression. 
%Nevertheless and in order to suppress this contribution, we further apply a  $M_T$ cut for the $e^+$ and the missing energy  $\slashed{E_T}$ system to ensure that there are decay products of the $W$ boson -- see Tab.~\ref{tabBckg100TeV_description} for a description of the cuts used in this analysis.

\end{enumerate}

For the signal generation, we use the  extension of the FeynRules package~\cite{Alloul:2013bka} for the minimal LR model  used  in Ref.~\cite{Roitgrund:2014zka} and expanded in Ref.~\cite{Nemevsek:2016enw}. The signal and background  events were generated at LO using Madgraph 5~\cite{Alwall:2014hca}, Pythia 6~\cite{Sjostrand:2006za} for hadronization, and Delphes 3~\cite{deFavereau:2013fsa} for detector simulation, using the JetFake module developed in \cite{Nemevsek:2016enw}.
The dominant sources of background are found to be $t \overline t W$, $t \overline t  (j) $ (with a jet faking a lepton) and $WWW(j)$, while $WZ(j)$, $t\overline t  Z$ and $Z/\gamma(j)$ (with charge flip and a jet faking a muon) are sub-dominant. The $j$ in the parenthesis means that we generated the corresponding background with one matched jet.
%
% $t \overline t  Z$,  $t \overline t W(j)$, $t \overline t  (j) $, $WZ  (j) $   and  $eej$. 
Tables \ref{tabBckg13TeV}, \ref{tabBckg28TeV} and \ref{tabBckg100TeV} show the cut flow (see below) for the main sources of background for this process, together with two signal benchmark points, for $13\mbox{ TeV}$, $28\mbox{ TeV}$ and $100\mbox{ TeV}$ respectively.  As already remarked,  we assume the left and right leptonic mixing matrix to satisfy $V_L=V_R^*$. Some of the backgrounds for this process were studied in Ref.~\cite{Sirunyan:2018mtv}. In our analysis, further sources of backgrounds are included mostly due to the charge misidentification probability that becomes more important at higher $p_T$.
We compare our $WZ$ and triple boson ($WWW$) backgrounds with the CMS estimates from Ref.~\cite{Sirunyan:2018mtv}.
In particular, we compare with the second last bin in the left panel of Figure A.3 from Ref.~\cite{Sirunyan:2018mtv}, which turns out to be closer to the kinematic region in our analysis, and find an agreement for $WZ$ and about half of the yield for the $WWW$ background. This difference is consistent with the 50\% uncertainty quoted for the estimate for triboson production. %in Ref.~\cite{Sirunyan:2018mtv}. 
%We try to reproduce the number of $WZ$ and triple boson ($WWW$) backgrounds in the second last bin in the left plot in Figure A.3 in Ref.~\cite{Sirunyan:2018mtv} which turns out to be closer to the kinematic region in our analysis, and find an agreement for $WZ$ background but about half of the number of $WWW$ background. We believe that this level of agreement is enough for our estimation of future experiment sensitivities in order of magnitude.
%
%%%
\begin{table} 
\centering
\tabcolsep=0.05cm
\begin{tabular}{ r | c  }%| c c c |}
\hline
 \hline 
          \centering Cut description  &   \\
\hline
\hline

    $e^+ e^+ \mu^-$, no $b$ jets and no additional leptons & signal selection \\
    $p_{T,e^+}^{lead}> 200  $  GeV, $p_{T,e^+}^{sub}> 100  $  GeV, $p_{T,\mu^-}^{lead}> 100$~GeV   & reduce all backgrounds  \\
    $\slashed{E}_T > 100$ GeV & reduce mostly $t \overline t (j)$ and  $Z/\gamma(j)$    \\
    $|m_{inv}(e^+ e^+) - 91.2|) > 10$ GeV & reduce mostly $WZ(j)$    \\
    $m_{T}(e^+_{sub} \slashed{E}_T)< 150\ {\rm GeV}$ &  \textcolor{blue}{} select   channel shown in Fig.~\ref{fig:KSlep} (right)   \\
    $m_T(e^+ e^+\mu^- \slashed{E}_T)> M_{W_R}/2$ &   reduce all backgrounds     \\
     \hline
\end{tabular}
\caption{Selection criteria used to reduced the SM background for 100 TeV. For 13 TeV and 28 TeV we apply the same cuts, excepting that $P^{lead}_{T,e^+} > 100\mbox{ GeV}$.}
\label{tabBckg100TeV_description}
\end{table}
%%%%

A  description of the selection criteria is shown in Table \ref{tabBckg100TeV_description}.
We first demand that each event contains exactly two positrons, one muon, and no b-tagged jets. Events with extra jets that are not b-tagged are retained. Secondly, we select events with high transverse momentum  $p_T$  for the leptons and large missing transverse energy  $\slashed{E_T}$ in order to reduce many of the backgrounds. Then we require the reconstructed invariant mass of the positron pair $m_{inv}(e^+ e^+)$ to be outside the Z boson mass peak, reducing the background coming from $Z\rightarrow e^+e^-$ when the electron charge is misidentified.

%%%%
 \begin{figure}\centering
     \includegraphics[scale=0.4]{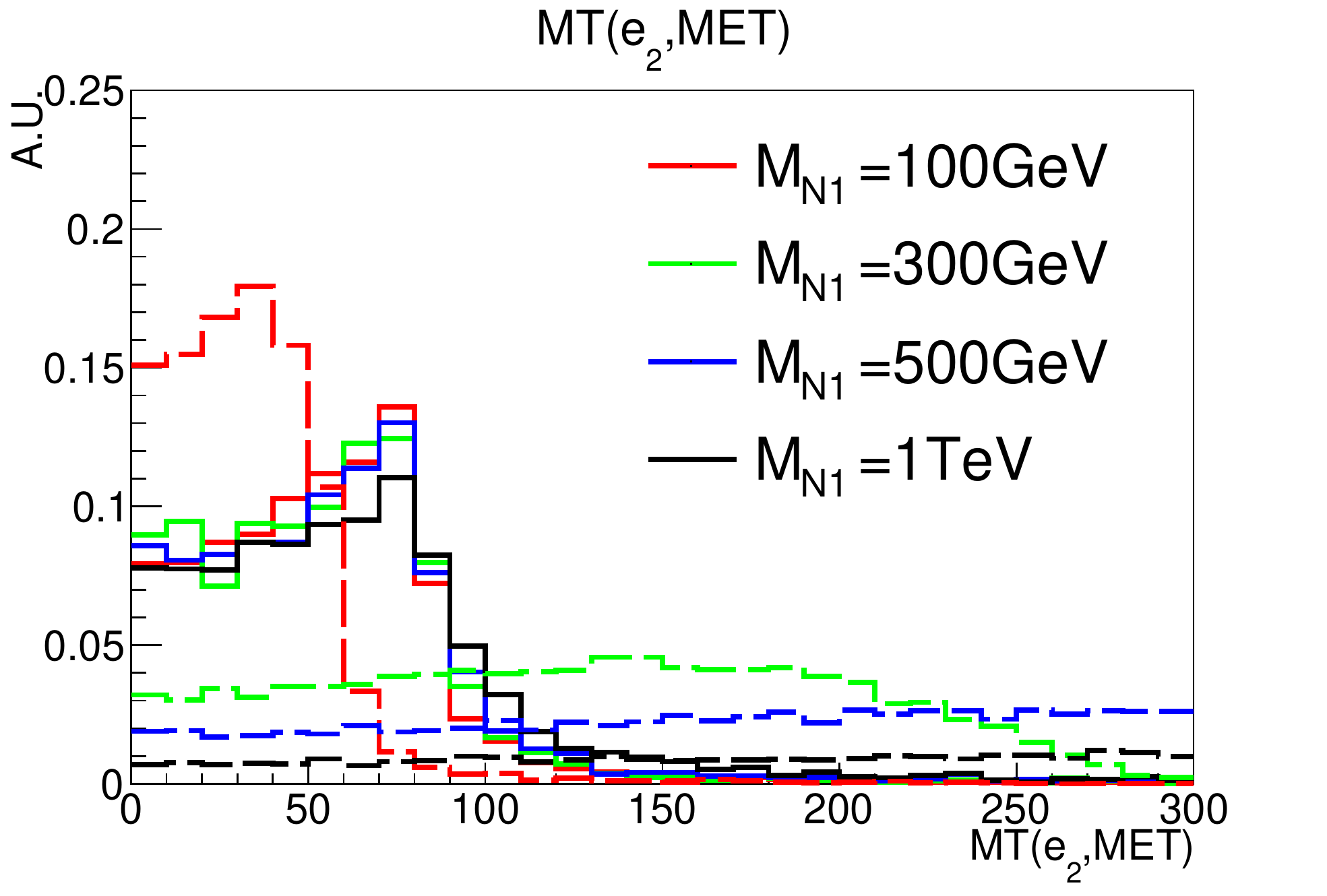}
  \caption{The distribution of the transverse mass of the subleading positron and missing $E_T$ at 100 TeV pp-collider with benchmark point $M_{W_R}=6$ TeV, $m_{N_2}=2$ TeV, $m_{N_3}=2.5$ TeV. Different colors represents different masses of $m_{N_1}$, the solid and dashed curves represent the events coming from the left and right diagram in Fig.~\ref{fig:KSlep} respectively.}
  \label{fig:MTe2met}
\end{figure}
%%%%%
The next cut in Table~\ref{tabBckg100TeV_description} is on the transverse mass of the positron and missing energy $m_T(e^+_{sub}~\slashed{E_T})$. We  enforce the reconstructed transverse mass of the sub-leading positron and missing energy to be less than 150 GeV. In principle, if the momentum of the leptons are exactly reconstructed, then $m_T(e^+_{sub}~\slashed{E_T})$ will not exceed the mass of the $W$ boson. However, due to smearing effects, the distribution of this transverse mass is broadened. This is why we choose the cut on this variable to be larger than the $W$ boson mass. In Fig.~\ref{fig:MTe2met} we show the $m_T(e^+_{sub}~\slashed{E_T})$ distribution for different masses of the heavy neutrino $N_1$ and $M_{W_R}=6$ TeV for a 100 TeV pp-collider. One can observe from these distributions that imposing a cut on this transverse mass can effectively discriminate the two diagrams in Fig.~\ref{fig:KSlep} if the mass of the $N_1$ is sufficiently large. As noted earlier, this discrimination can be achieved because the positron from the heavy neutrino decay comes mostly from an on-shell $W$ boson decay in the process in Eq.~(\ref{posi2}). The signal efficiencies are different for the two processes in Fig.~\ref{fig:KSlep}. In Fig.~\ref{fig:sigeff} we show the signal efficiency for each channel individually as well as the averaged efficiencies with different relative strength of the two channels characterized by the parameter $r$ defined below:
\begin{eqnarray}
r \equiv \frac{Br(N_1\to e^+(W^-\to \mu^-\bar{\nu_\mu}) )}{Br(N_1\to \mu^-(W^+\to e^+\nu_e))}.\label{eq:r}
\end{eqnarray}
{As one can see from the left plot in Fig.~\ref{fig:sigeff}, the efficiency of the channel shown in Eq.~\eqref{posi2}  decreases as the mass of the $N_1$ increases. This is mainly  due to the  cut  \mbox{$m_T(e^+_{sub} \slashed{E}_T) < 150$ GeV} shown in Table \ref{tabBckg100TeV_description}, which helps to discriminate between the  two channels shown in Eqs.~\eqref{posi1}  and \eqref{posi2}.}  

 \begin{figure}\centering
     \includegraphics[scale=0.3]{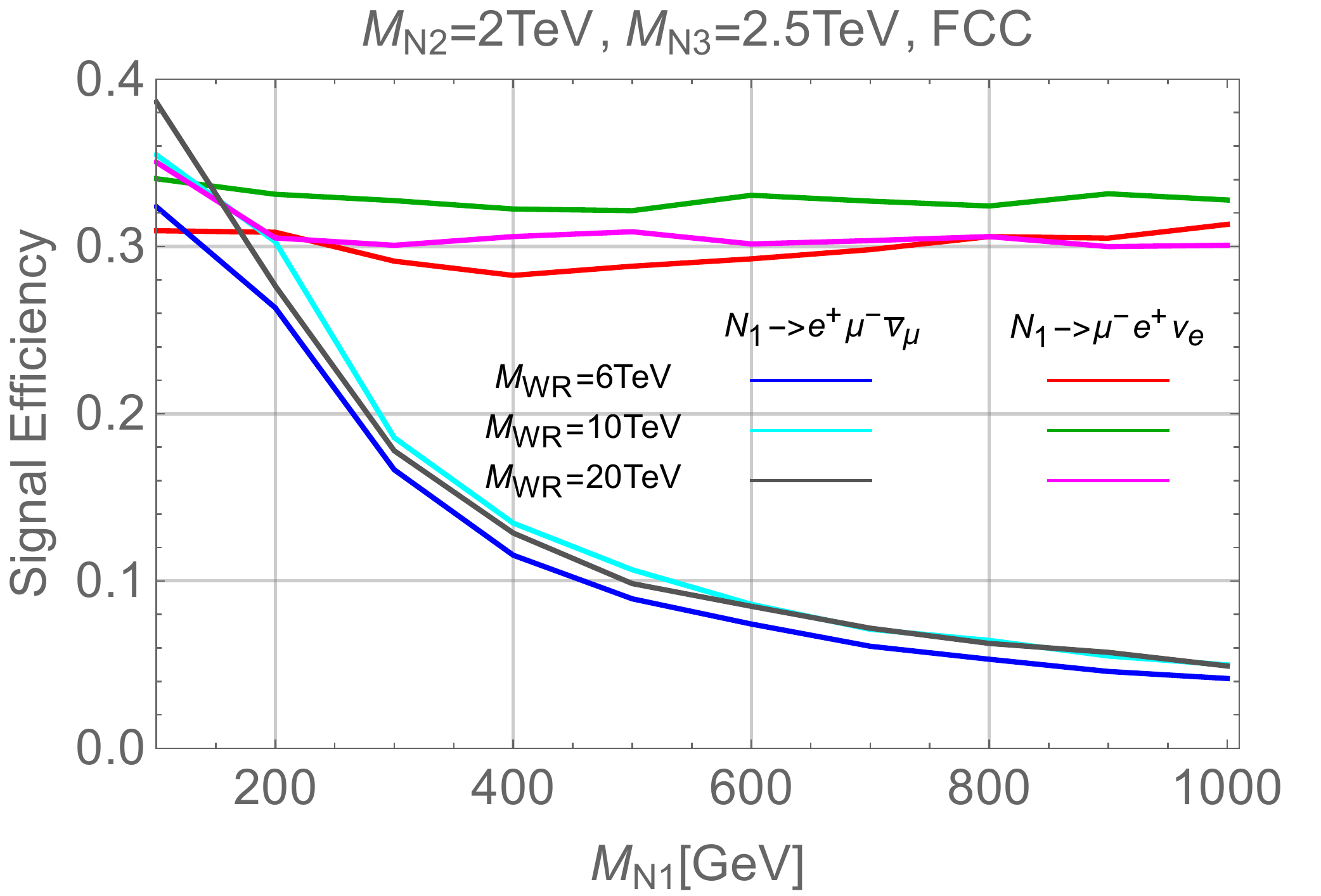}   
      \includegraphics[scale=0.3]{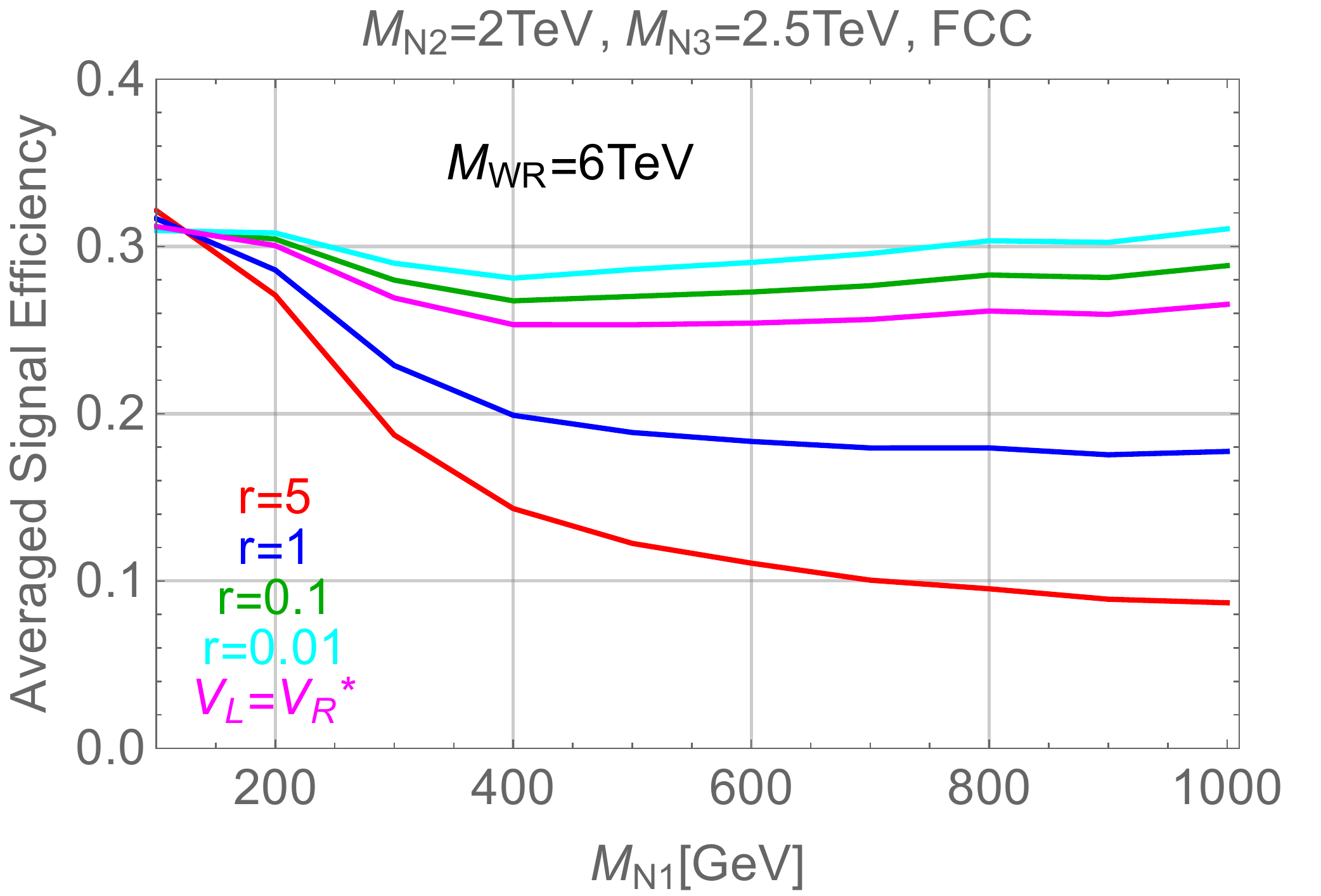} 
  \caption{The signal efficiencies for the benchmark point: $m_{N_2}=2$ TeV, $m_{N_3}=2.5$ TeV. The left plot shows the signal efficiencies for each channel with different mass of $W_R$ and $N_1$. The right plot shows the signal efficiencies for different relative strengths $r$ of the two channels (defined in Eq.~\ref{eq:r}), with $M_{W_R}=6$ TeV.}
  \label{fig:sigeff}
\end{figure}

{The last selection in Tab.~\ref{tabBckg100TeV_description} is a  cut on  the transverse mass of the $e^+e^+\mu^-\slashed{E_T}$ system, since for an on-shell $W_R$ boson,  the transverse mass distribution is peaked at $M_{W_R}$ (see Fig.~\ref{fig:MTWR}), where the SM backgrounds give a   negligible contribution.   The rejection of the backgrounds  was  effectively  achieved by using   the cut $m_T(e^+e^+\mu^-\slashed{E_T})> M_{W_R}/2$ shown in Table~\ref{tabBckg100TeV_description}. In this way, most of the signal events are kept while a significant portion of the backgrounds is rejected. }
Furthermore, this cut also guarantees  that  the SM  backgrounds become even more suppressed when searching for a $W_R$ boson with higher mass.
%{\color{magenta} I did not see a discussion of how these distributions inform the choice of cuts in Table 1}
%the $p_T$ of the lepton coming from the $W_R$ decay would be peaked at $p_T=\frac{M_{W_R}}{2}$.

\begin{figure}\centering
     \includegraphics[scale=0.4]{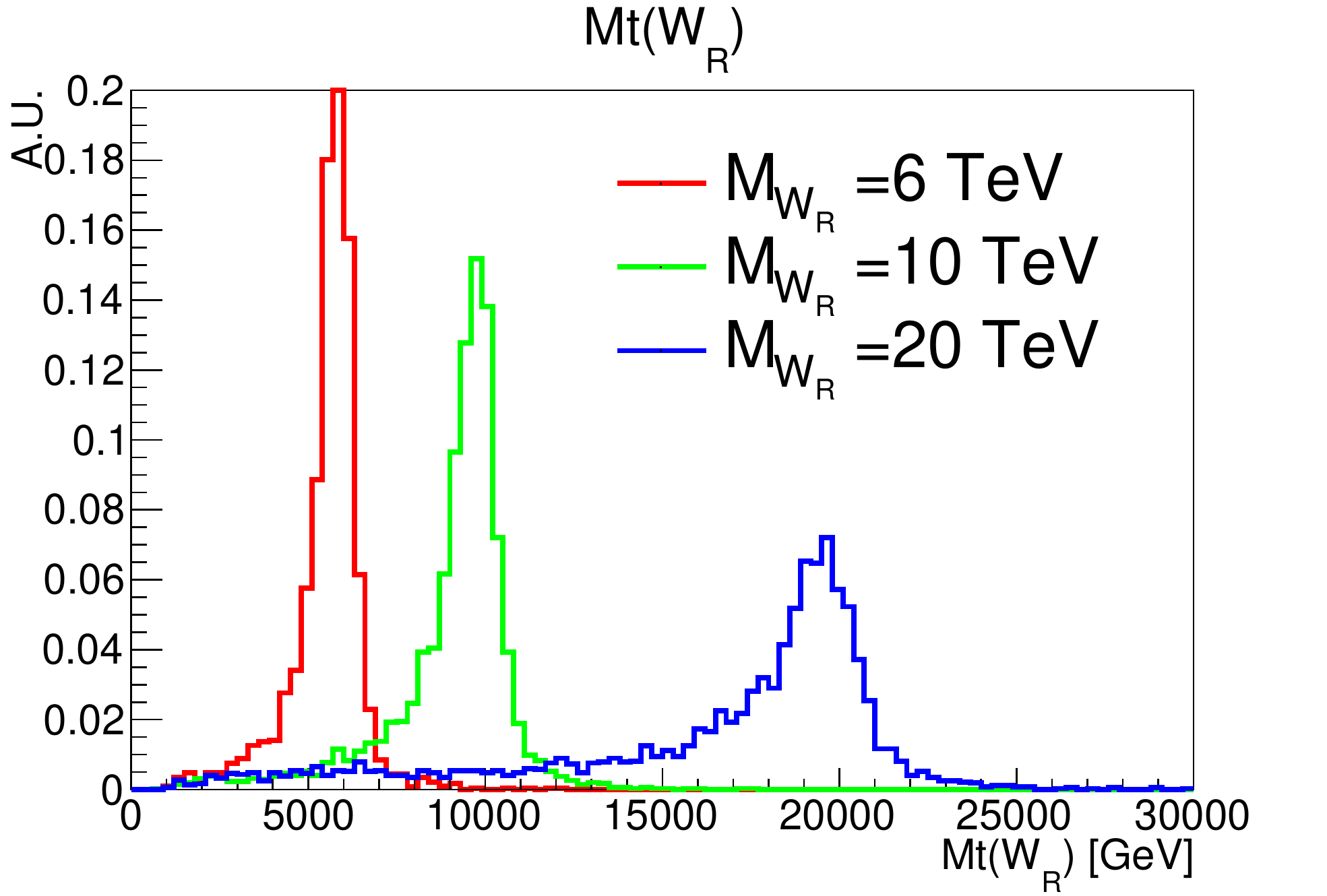}
  \caption{The distribution of the reconstructed transverse mass of $W_R$ at a 100 TeV pp-collider, with benchmark point $m_{W_N}=300$ GeV, $m_{N_2}=2$ TeV, $m_{N_3}=2.5$ TeV. Different colors represents different masses of $W_R$.}
  \label{fig:MTWR}
\end{figure}

For the charge flip probability, we take the current ATLAS performance in Ref.~\cite{Aaboud:2017qph}, which parameterizes the flip probability $P$ as the product of functions of $\eta$ and $P_T$: \mbox{$P = f(\eta)\times\sigma$($P_T)$}. Also we assume that $\sigma$($P_T$) for $P_T>$ 400~GeV keeps the same value as that in the bin (200,400) GeV.
For our analysis of the 100 TeV collider reach, we use the same charge flip probability, as we are not aware of a more realistic estimation having appeared in the literature to date.

%since no estimation of this information is given in the literature so far.} We use the  {\color{magenta} It seems we use this for 100 TeV even though we don't know this will be correct...some mention of this fact should be made} 
%%
 \begin{figure}\centering
     \includegraphics[scale=0.26]{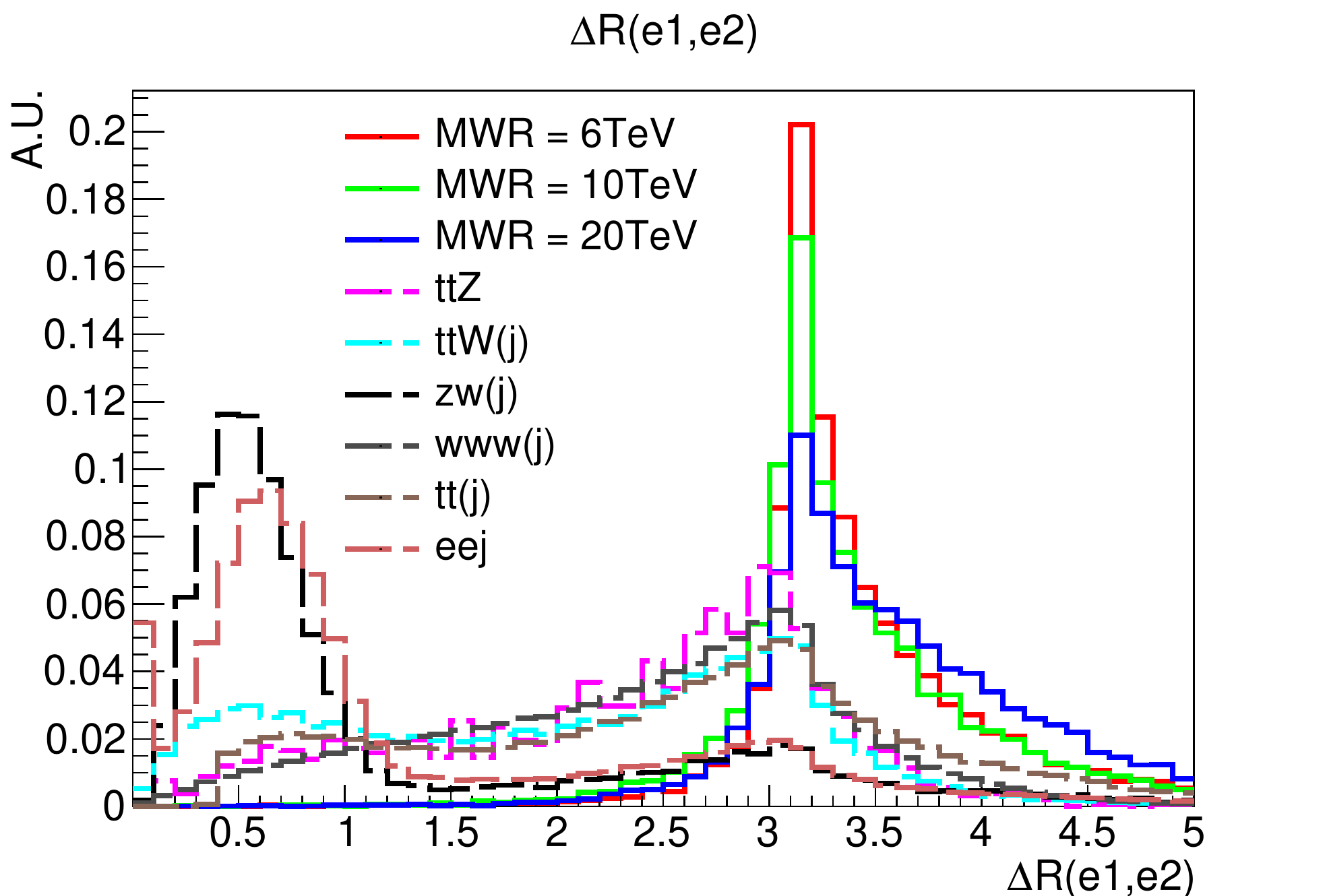}   
      \includegraphics[scale=0.26]{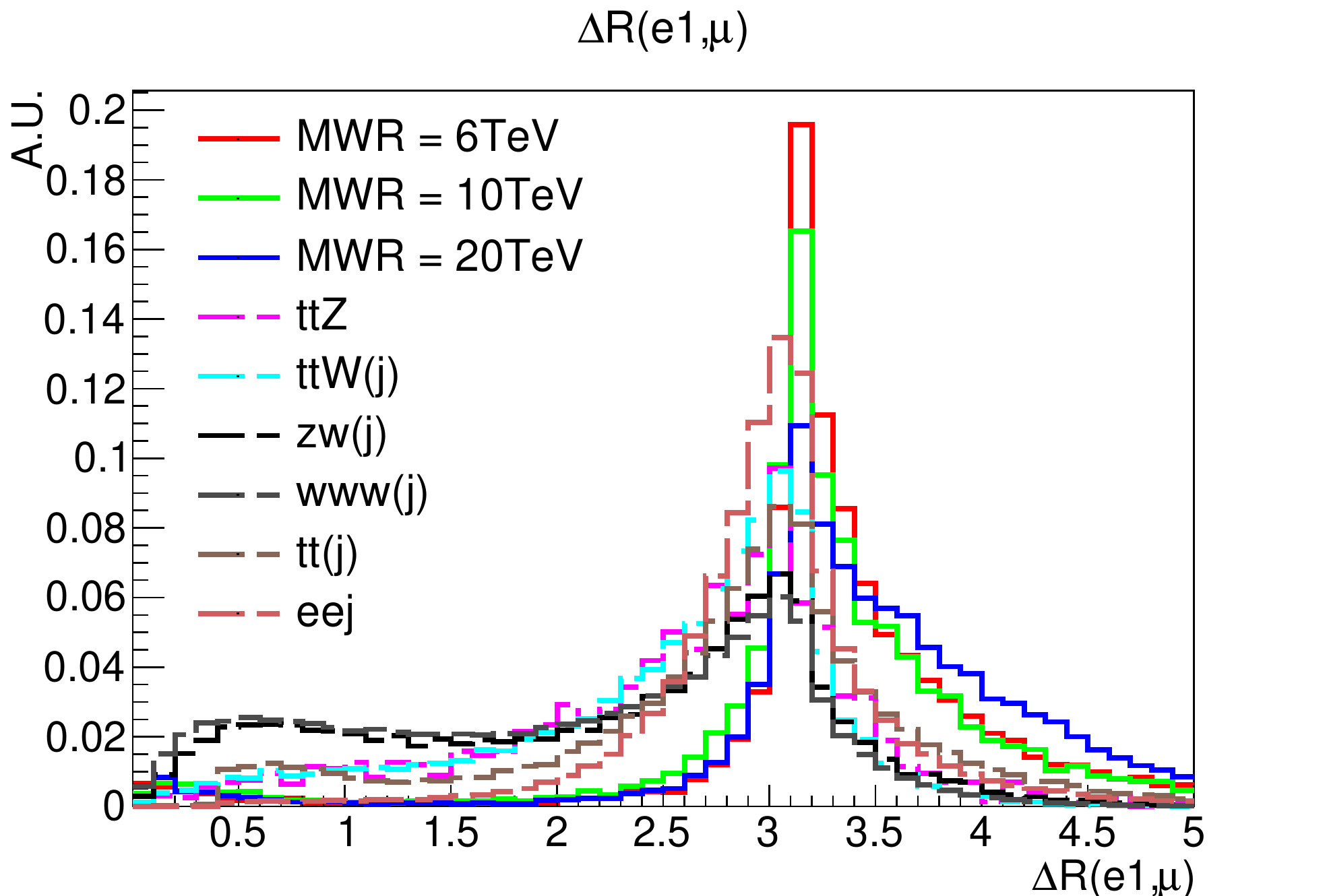} 
       \includegraphics[scale=0.26]{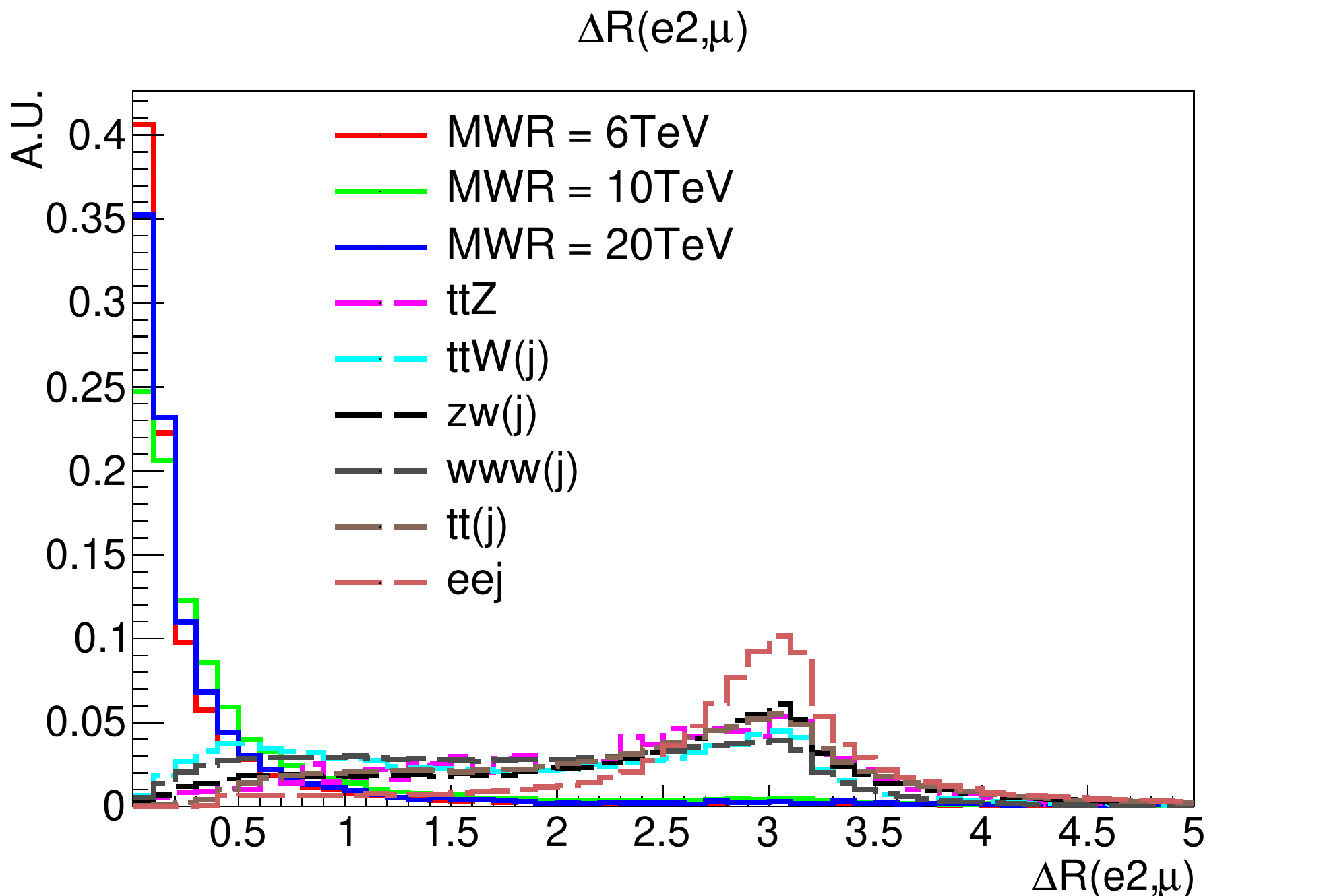}  
  \caption{$\Delta R$ for the decay products of the process shown in Fig.~\ref{fig:KSlep}. $e_1$ and $e_2$ refers to the harder and softer positrons sorted  by transverse momentum.  We assume $m_N = 100$ GeV and plot the signal for different values of $M_{W_R}$.    }
  \label{fig:DR}
\end{figure}

Finally, a comment on a possible future refinement that can be made in the analysis and that exploits a particular kinematic feature of the signal. For HN  masses much smaller than the $W_R$ boson mass, the decay products coming from the HN are merged and form what it has been coined as a \emph{lepton jet}~\cite{Izaguirre:2015pga}.  In Fig. \ref{fig:DR}, we show  $\Delta R \equiv\sqrt{\Delta \phi^2+\Delta \eta^2}$  for the decay products of the process shown in Fig.~\ref{fig:KSlep}.  Since $\Delta R(\mu e_2)$ for the signal is  peaked at  smaller values  (around $\Delta R \sim 5 \times  10^{-2}$ for $m_N= 100$ GeV and for $M_{W_R}=6,10,20$ TeV) than $\Delta R(\mu e_1)$ and $\Delta R(e_1 e_2)$, we see that the positron with smaller energy is mainly coming from the HN decay. This kinematic variable has been previously proposed in Ref.~\cite{Chen:2013fna}. Although we did not need to use this kinematic feature in our analysis, it is worth to keep it in mind, since it is one of the more distinctive topological features of the signal in the mass range considered in this work and it may be used to reject further backgrounds.

%%%%%%
\subsection{Sensitivity at the LHC }

 \begin{figure}\centering
     \includegraphics[scale=0.6]{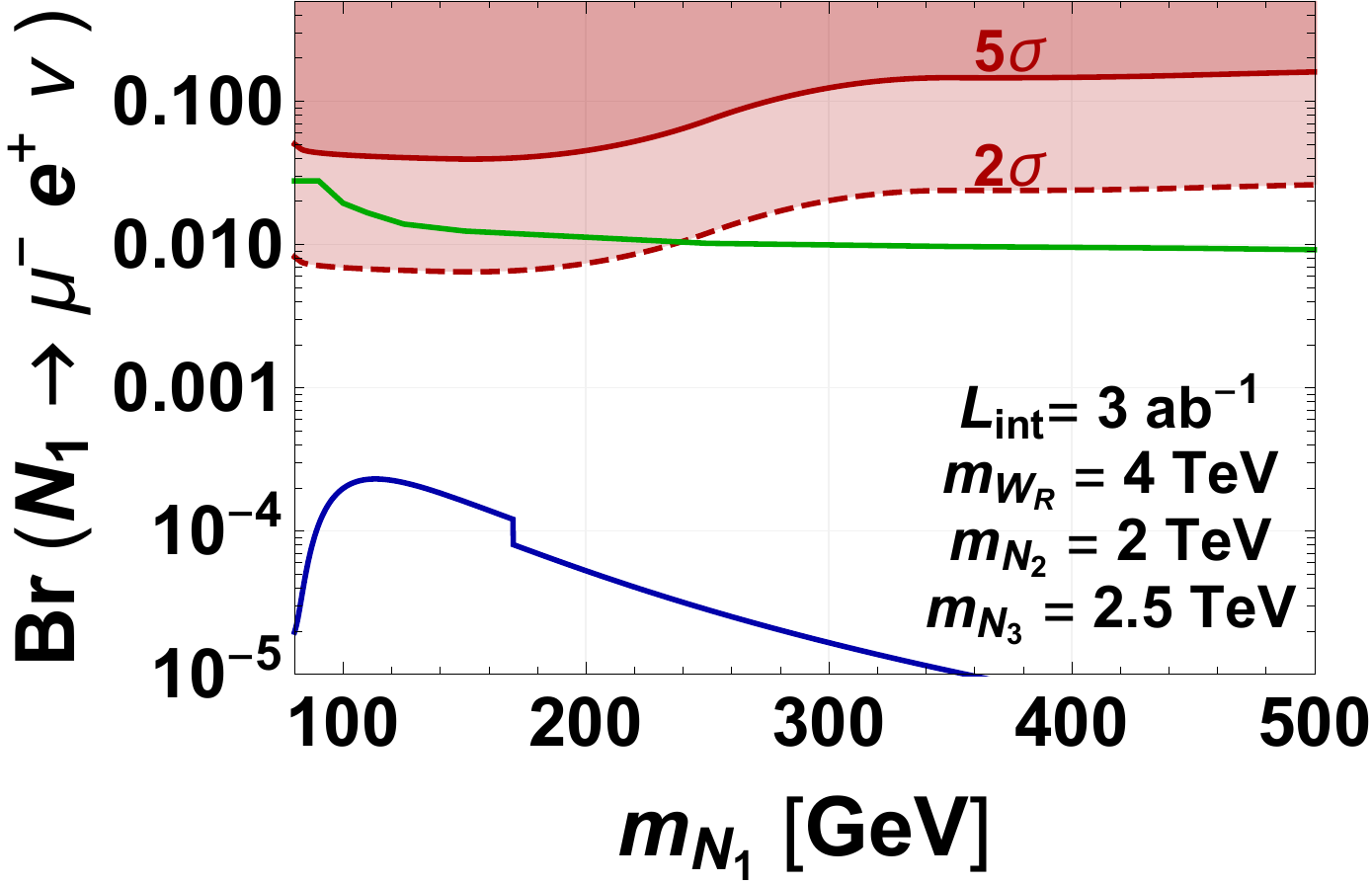}  
  \caption{LHC reach to the branching ratio of the purely leptonic decays of the heavy neutrino. The blue (green) line denotes the branching ratio within the minimal (non-minimal) LR model and the shadowed thick(dashed) regions show the reach at $5\sigma(2\sigma)$, for an integrated luminosity of $L_{int} = 3 \text{ab}^{-1}$ and center of mass energy $\sqrt{s}= 13$ TeV. We assume $V_L=V_R^*$ and the  upper limit on light neutrino masses of $m_{\nu}=0.5$ eV~\cite{Nakamura:2010zzi}. For the non-minimal model we have   set    ${\cal R} = U_R =I$ and $\mu = 10^{-4}$ GeV.    }
  \label{fig:br2lep2_LHC}
\end{figure}
\begin{table}
%\begin{table} 
\centering
\tabcolsep=0.05cm
\begin{tabular}{ r | c c c c c c | c  c c c |}%| c c |}
\hline
\hline
 & & \multicolumn{4}{c}{Backgrounds}& & \multicolumn{2}{c}{Signal} \\
 \hline 
          $\sqrt{s}=$13TeV&  $t \overline t  Z$ & $t \overline t W$ & $t \overline t ( j) $ & $WZ  (j) $   & $3W$ & $Z/\gamma(j)$ &$m_N (100$ GeV) & $m_N( 500$ GeV) \\
\hline
\hline
$e^+e^+\mu^-$ (b-veto)            & 11.8  & 74.9 & 23058 & 24.8 & 6.71 & 901 & 1293 & 371  \\
$P_T\mbox{ cuts}$                 & 0.325 & 3.75 & 216 & 0.215  & 2.33 & 5.31 & 825 & 253 \\
$\slashed{E}_T\ {\rm GeV}$   & 0.158 & 1.85 & 117 & 0.0761 & 1.06 & 0.0911 & 646 & 188 \\
$m_{inv}(e^+\ e^+)$               & 0.155 & 1.82 & 113 & 0.0761 & 1.05 & 0 & 646 & 188 \\
$m_{T}(e^+_{sub}\ \slashed{E}_T)$ & 0.0582& 0.743 & 48.4 & 0.0277 & 0.491 & 0 & 622 & 176 \\
$m_T(e^+ e^+\mu^- \slashed{E}_T)$ & 0     & $7.82\times 10^{-3}$ & 0 & 0 & 0.0169 & 0 & 597 & 158
   \\ \hline
\end{tabular}
\caption{SM background processes at 13 TeV and $3 \text{ ab}^{-1}$ for the trilepton signal $e^+ e^+ \mu^- \nu$ and $M_{W_R}= 4$ TeV, for two benchmark values of the heavy neutrino masses assuming $(\Theta_L)_{\mu N}=(\Theta)_{e N}=\frac{1}{\sqrt{2}}$.
Backgrounds ending with $(j)$ were simulated with one matched jet.
The charge misidentification probability has been taken from current ATLAS result from Ref.~\cite{Aaboud:2017qph}.
%The NLO K-factor for backgrounds are taken from Ref.~\cite{Mangano:2016jyj}
%(*) This backgrounds were obtained using the default Madgraph generation  card.
}
\label{tabBckg13TeV}
%\end{table}
\end{table}

The LHC sensitivity to the branching ratio into the purely leptonic channel is shown in Fig.~\ref{fig:br2lep2_LHC}, with the corresponding cut flow for representative two signal points given in Table~\ref{tabBckg13TeV}\footnote{In this analysis, we studied the sensitivity of the trilepton signal for masses $m_N \gtrsim 80$ GeV. A similar analysis can be carried out for smaller masses, although in this case the $N$ could decay with a displaced vertex, as previously studied in Refs.\cite{Nemevsek:2011hz, Nemevsek:2018bbt, Helo:2013esa, Cottin:2018kmq, Cottin:2018nms}}. {In our calculations we assumed $V_L=V_R^*$  and set ${\cal R} = U_R =I$, $\mu = 10^{-4}$ GeV for the non-minimal model. This choice of parameters  corresponds to $r\sim 1$ in both scenarios}.\footnote{{ In the inverse see saw scenario different values of $r\sim [0,1]$ are possible even if we assume $V_L=V_R^*$. This is because the lepton number violating process  $p p \rightarrow W_R^+ \rightarrow e^+ e^+ \mu^- \bar{\nu}_\mu$ in Eq. (\ref{posi2}) might be suppressed with respect to the  lepton number conserving process $p p \rightarrow W_R^+ \rightarrow e^+ \mu^- e^+  \nu_e$ in Eq. (\ref{posi1}) for $\mu < \Gamma_N $, being $\Gamma_N$ the decay width of the heavy neutrino (see the discussion in Refs.~\cite{Anamiati:2016uxp, Das:2017hmg}). }} { As expected the maximum reach is obtained for $M_{W_R}= 4$ TeV, for a wide range of the heavy neutrino mass, with the branching ratio reach extending down to $\sim 10^{-2}$ at $2\sigma$ significance. 
%We plot the expected reach for $M_{W_R}= 4$ TeV that is around  $10^{-2}$ at $2\sigma$ respectively.  
The solid blue and green curves show the expected branching ratios in the mLRSM and non-minimal models, respectively. It is clear that for the mLRSM, the branching ratio lies well below the sensitivity the HL-LHC ($L_{int}=3ab^{-1}$).  This suppression with respect to the LHC reach follows from the small values of $M_D/M_N$ needed for consistency with the observed small scale of the light neutrino masses that imply, in turn,  small values of the heavy-light neutrino mixing angle. This suppression is not necessarily true in an inverse seesaw scenario. In the latter case, the smallness of the neutrino masses can be attributed to small values of the $\mu$ parameter (see Eq. (\ref{numass3})) without requiring the exceedingly small values of $M_D/M_N$\cite{Anamiati:2016uxp, Das:2016akd,Das:2017hmg}. This feature can 
be seen explicitly in Fig.~\ref{fig:br2lep2_LHC}, where we have obtained the green curve by choosing $\mu = 10^{-4}$ GeV. Therefore, we can conclude that a signal of the purely leptonic channel at the LHC would be an indication of a non-minimal realization of the LR symmetry. } 
%\textcolor{blue}{A word of caution must be placed here, strictly speaking the exclusion  curves shown in Fig.~\ref{fig:br2lep2_LHC} apply to the minimal model only with $V_R^*=V_L$, namely $r\sim1$. However, from Fig.~\ref{fig:sigeff} we see that since for the minimal model $r\sim 1$   and   $r < 1$  (see the discussion in Ref.~\cite{Anamiati:2016uxp}) for the inverse see-saw scenario, the difference in the efficiency  between  the two scenarios does not vary significantly. }

%
\subsection{Sensitivity at the HE-LHC}

 \begin{figure}\centering
     \includegraphics[scale=0.35]{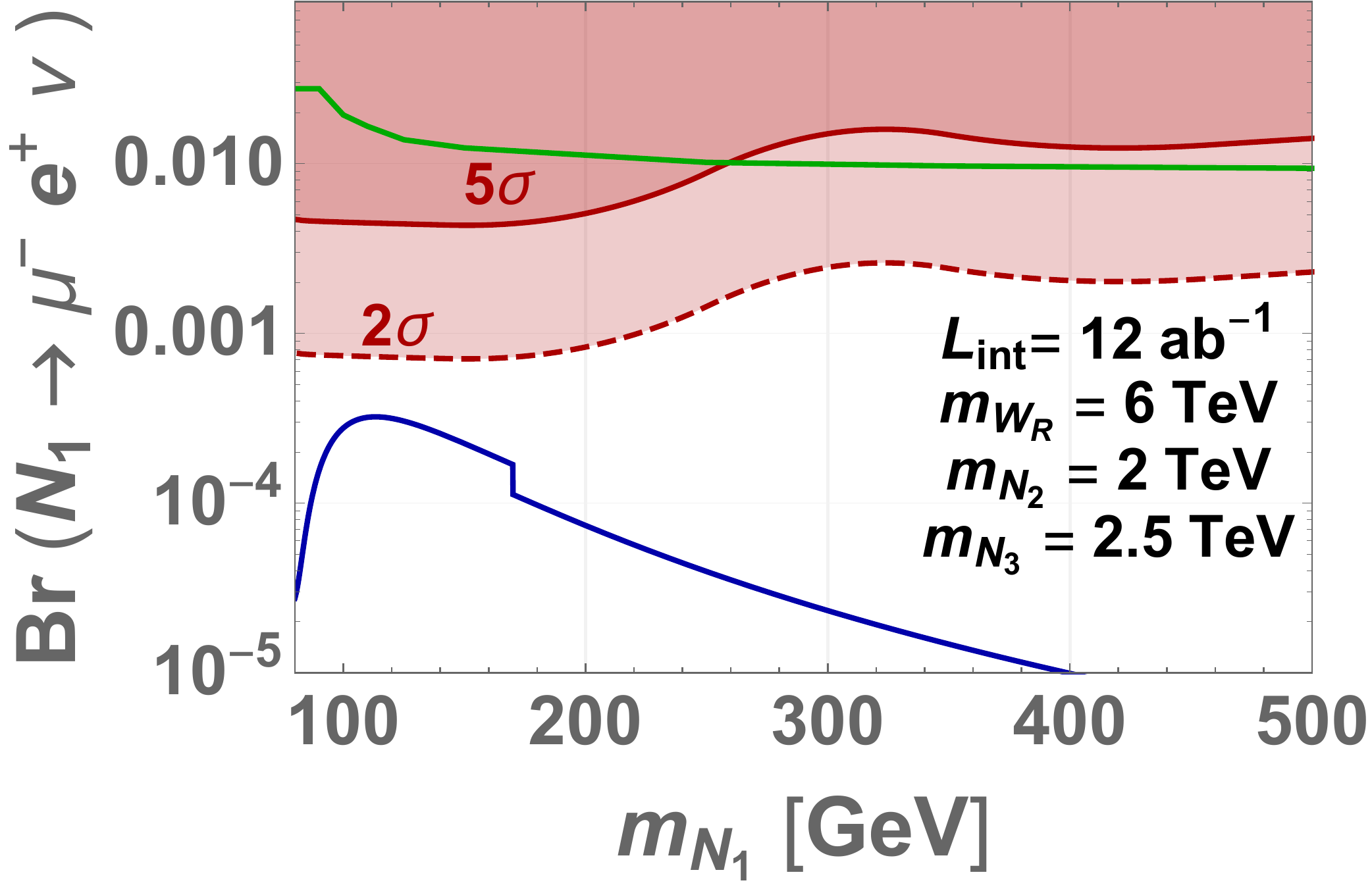}   \includegraphics[scale=0.35]{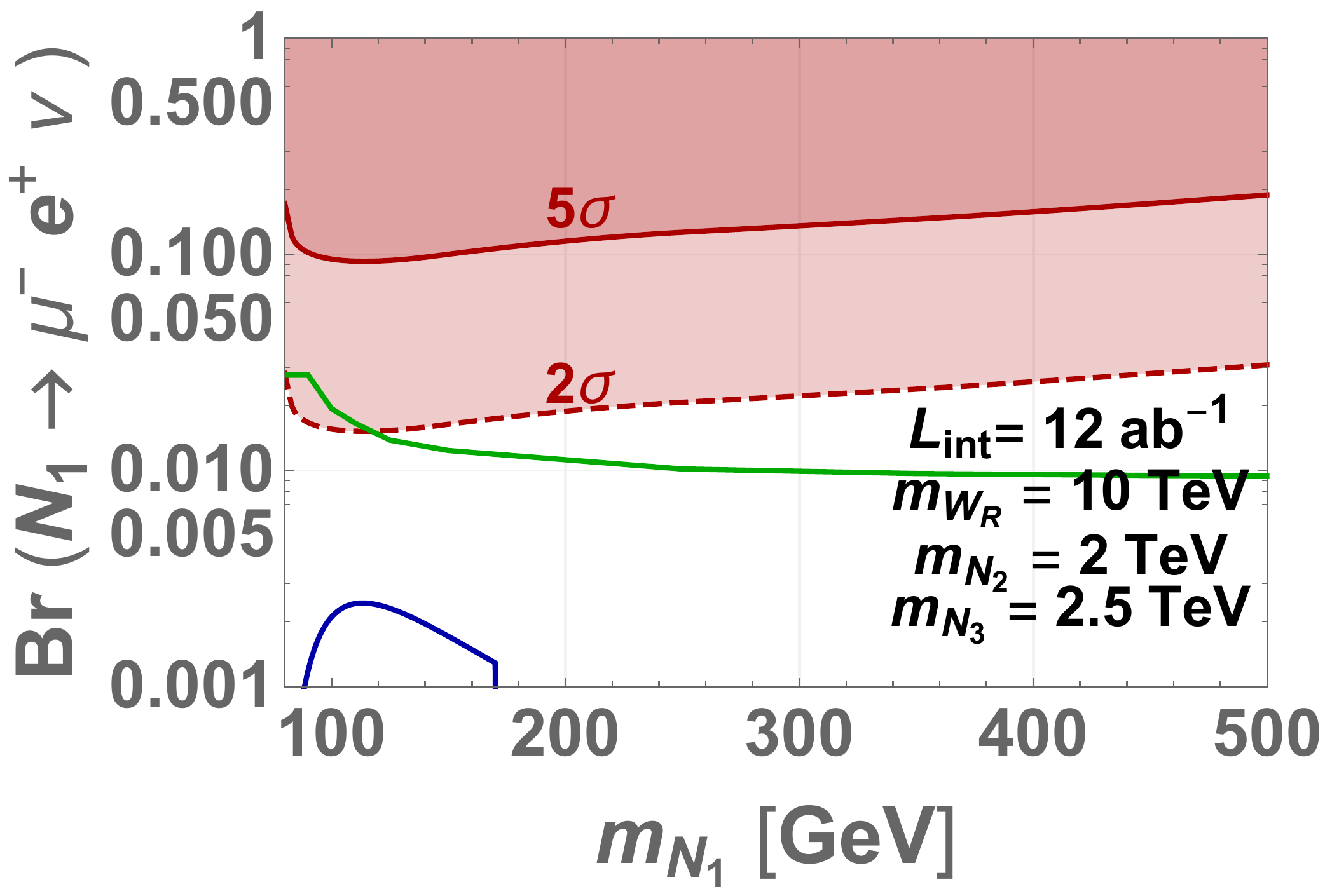}
  \caption{HE-LHC reach to the branching ratio of the purely leptonic decays of the heavy neutrino. The blue (green) line denotes the branching ratio within the minimal (non-minimal) LR model and the shadowed thick(dashed) regions show the reach at $5\sigma(2\sigma)$, for an integrated luminosity of $L_{int} = 12 \text{ab}^{-1}$ and center of mass energy $\sqrt{s}= 28$ TeV.  We assume $V_L=V_R^*$ and the  upper limit on light neutrino masses of $m_{\nu}=0.5$ eV~\cite{Nakamura:2010zzi}. For the non-minimal model we have   set    ${\cal R} = U_R =I$ and $\mu = 10^{-4}$ GeV.  }
  \label{fig:br2lep2_HE_LHC}
\end{figure}
\begin{table}
\centering
\tabcolsep=0.05cm
\begin{tabular}{ r | c c c c c c | c  c c c |}%| c c |}
\hline
\hline
 & & \multicolumn{4}{c}{Backgrounds}& & \multicolumn{2}{c}{Signal} \\
 \hline 
          $\sqrt{s}=$28TeV&  $t \overline t  Z$ & $t \overline t W$ & $t \overline t ( j) $ & $WZ  (j) $   & $3W$ & $Z/\gamma(j)$ &$m_N (100$ GeV) & $m_N( 500$ GeV) \\
\hline
\hline
$e^+e^+\mu^-$ (b-veto)            & 286  & 882 & 615657 & 440 & 56.6 & 3139 & 8766 & 3208  \\
$P_T\mbox{ cuts}$                 & 9.93 & 60.0 & 8791 & 6.3 & 22.9 & 37.9 & 7000 & 2474 \\
$\slashed{E}_T\ {\rm GeV}$   & 5.18 & 34.1 & 5115 & 2.29 & 12.2 & 2.33 & 6037 & 2092 \\
$m_{inv}(e^+\ e^+)$               & 5.00 & 33.6 & 4986 & 2.29 & 12.2 & 0.608 & 6037 & 2092 \\
$m_{T}(e^+_{sub}\ \slashed{E}_T)$ & 2.14 & 13.7 & 2297 & 0.497 & 5.83 & 0 & 5678 & 1883 \\
$m_T(e^+ e^+\mu^- \slashed{E}_T)$ & 0   & 0.028 & 3.00 & 0 & 0.13 & 0 & 5555 & 1800
\\ \hline
\end{tabular}
\caption{SM background processes at 28 TeV and $12 \text{ ab}^{-1}$ for the trilepton signal $e^+ e^+ \mu^- \nu$ and $M_{W_R}= 6$ TeV, for two benchmark values of the heavy neutrino masses assuming $(\Theta_L)_{\mu N}=(\Theta)_{e N}=\frac{1}{\sqrt{2}}$.
Backgrounds ending with $(j)$ were simulated with one matched jet.
The charge misidentification probability has been taken from current ATLAS result from Ref.~\cite{Aaboud:2017qph}.
%The NLO K-factor for backgrounds are taken from Ref.~\cite{Mangano:2016jyj}
%(*) This backgrounds were obtained using the default Madgraph generation  card.
}
\label{tabBckg28TeV}
\end{table}

We now consider the reach of the proposed energy upgrade of the  LHC to  $28$ TeV center of mass energy (HE-LHC).  In Fig.~\ref{fig:br2lep2_HE_LHC} we show the expected reach at the HE-LHC for two values of the $W_R$ boson mass. We see from the figure that heavy light mixing can be excluded at  $2\sigma$ for heavy-light mixing of the order of $10^{-3}$ for $M_{W_R}=6$ TeV and of the order of $10^{-2}$ for $M_{W_R}=10$ TeV. As in the case of the high luminosity LHC (HL-LHC), the minimal LR scenario is not expected to give any observable signal in the purely leptonic channel, so a positive signal would point to a non-minimal realization of the LR model.  Finally and for the purpose of illustration, in Table~\ref{tabBckg28TeV} we show the cut flow for the main background processes and two benchmark points for the signal.

\subsection{Sensitivity at the 100 TeV $pp$ collider }

We now turn to the reach of a 100 TeV $pp$ collider, as shown in Fig. \ref{fig:br2lep2_FCC}. The corresponding cut flow is given in Table \ref{tabBckg100TeV}. 
%we show the cut flow for the main background processes to the leptonic process $ p p \rightarrow e^+ N \rightarrow e^+ N \rightarrow e^+  \mu^- e^+ \nu $ and for two benchmark values for the HN masses. 

%%

%
%\begin{sidewaystable}
\begin{table} 
\centering
\tabcolsep=0.05cm
\small{%%
\begin{tabular}{ r | c c c c c c | c  c c c |}%| c c |}
\hline
\hline
 & & \multicolumn{4}{c}{Backgrounds}& & \multicolumn{2}{c}{Signal} \\
 \hline 
          $\sqrt{s}=$100TeV&  $t \overline t  Z$ & $t \overline t W$ & $t \overline t ( j) $ & $WZ  (j) $   & $3W(j)$ & $Z/\gamma(j)$ &$m_N (100$ GeV) & $m_N( 500$ GeV) \\
\hline
\hline

$e^+e^+\mu^-$ (b-veto)   &199 & 1.1K  &  1.2K  &  9K & 735 & 1.1K & 1.9M &1.8M \\
$P_T\mbox{ cuts}$  &  18.7 & 387 & 226  & 2.4K  &  254 & 244  & 1.34M  &  1.30M \\
    $\slashed{E}_T$ & 12.6  &  312 &  138  & 1.1K   & 165 &    18.7 & 1.1M  & 1M \\
    $m_{inv}(e^+\ e^+) {\rm cuts}$ &   12.1   &   311  &    136   &   122   & 164  &  5.19  &  1.1M &  1M\\
    $m_{T}(e^+_{sub}\ \slashed{E}_T)\ $ &   4.42   &   116  &    65.1   &   22   &  85.9 &  0.344  &  1.1M &  0.99M\\
    $m_{inv}(e^+ e^+\mu^- \slashed{E}_T)$ &   0.126   &   7.60  &    5.82    &  0.336   &  9.72 & 0.0275  &  1M & 0.97M \\
    %$m_{inv}(e^+ e^+\mu^- \slashed{E}_T)>5\ {\rm TeV}$ &   0   &   %0.918  &    1.11    &   0   &    1.94 & 0 &  --- & --- \\
    %$m_{inv}(e^+ e^+\mu^- \slashed{E}_T)>10\ {\rm TeV}$ &   0   &   2.62$\times 10^{-2}$  &    6.16$\times 10^{-2}$    &   0   & 8.05$\times 10^{-2}$  &  0 &  --- & --- 
   \\ \hline
\end{tabular}
\caption{SM background processes at 100 TeV and $30 \text{ ab}^{-1}$ for the trilepton signal $e^+ e^+ \mu^- \nu$ and $M_{W_R}= 6$ TeV, for two benchmark values of the heavy Neutrino masses assuming $(\Theta_L)_{\mu N}=(\Theta)_{e N}= \frac{1}{\sqrt{2}}$. Backgrounds ending with $(j)$ were simulated with one matched jet.
The charge misidentification probability has been taken from current ATLAS result from Ref.~\cite{Aaboud:2017qph}. The jet to lepton fake rates for $t\bar t(j)$ and $Z/\gamma(j)$ have been taken as $10^{-4}$ universally. The NLO K-factor for backgrounds are taken from Ref.~\cite{Mangano:2016jyj}
%(*) This backgrounds were obtained using the default Madgraph generation  card.
}
\label{tabBckg100TeV}
}%%
\end{table}
%\end{sidewaystable}
{In contrast to the situations for the HL-LHC and HE-LHC, a 100 TeV $pp$ collider could observe the trilepton channel for a sufficiently light $W_R$. As one increases the $W_R$ mass, the range of HN masses for which the trilepton channel is accessible decreases. 
From the left panel of Fig. \ref{fig:br2lep2_FCC}, we see that  for $M_{W_R} = 6$ TeV  the purely leptonic signal can be discovered(excluded) with $5\sigma(2\sigma)$ sensitivity  for heavy neutrino masses below  300 GeV(450 GeV). It is, thus, possible that one might discover the $W_R$ boson at the LHC (with $M_{W_R}< 6$ TeV) using the two lepton and two jets channel~\cite{Nemevsek:2018bbt} yet require a 100 TeV $pp$ collider to probe the HN mass generation by measuring the trilepton signal.  For heavier $W_R$ masses, both discovery of the RH $W$ boson and testing the mLRSM Yukawa sector would require a next generation collider. For $M_{W_R} = 10$ TeV, for example, the purely leptonic signal can be discovered(excluded) with $5\sigma(2\sigma)$ sensitivity  for heavy neutrino masses below  260 GeV(400 GeV). Finally, for  $M_{W_R} = 20$ TeV heavy neutrino masses up to 200 GeV can be excluded at $2\sigma $.}

%\textcolor{red}{In Fig. \ref{fig:br2lep2_FCC} sensitivity limits start at $m_{N_1} = 20 GeV$ [X]. However for  masses $m_{N_1} \lesssim 50 GeV$ the heavy neutrino will start to decay from a displaced vertex. We din't take into account this effect in our analysis. Therefore our limits for haevy neutrino masses $m_{N_1} \lesssim 50 GeV$ are optimistics. For a detailed analysis of displaced vertex in LR models, see [X].s }

 \begin{figure}\centering
     \includegraphics[scale=0.25]{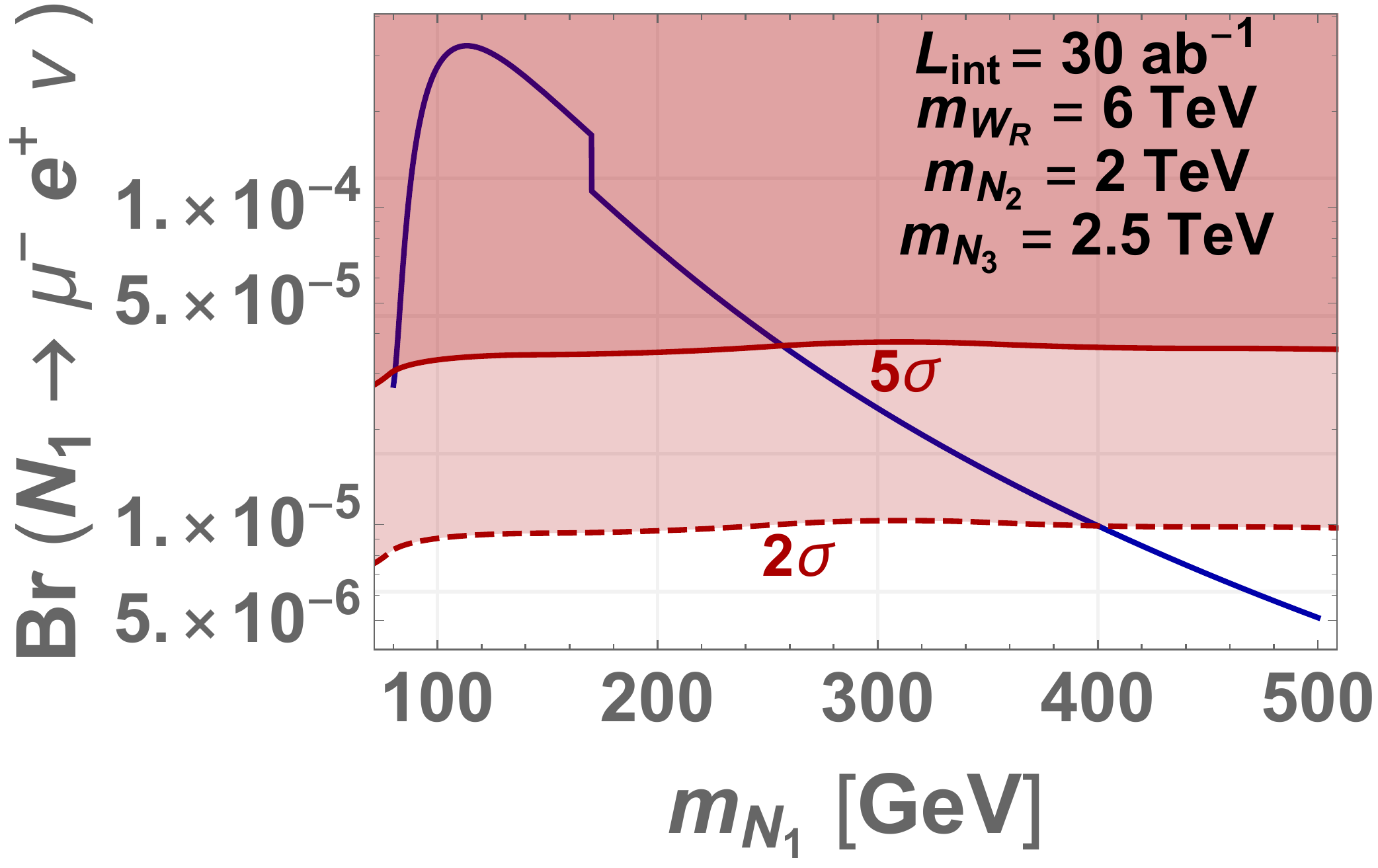}   \includegraphics[scale=0.25]{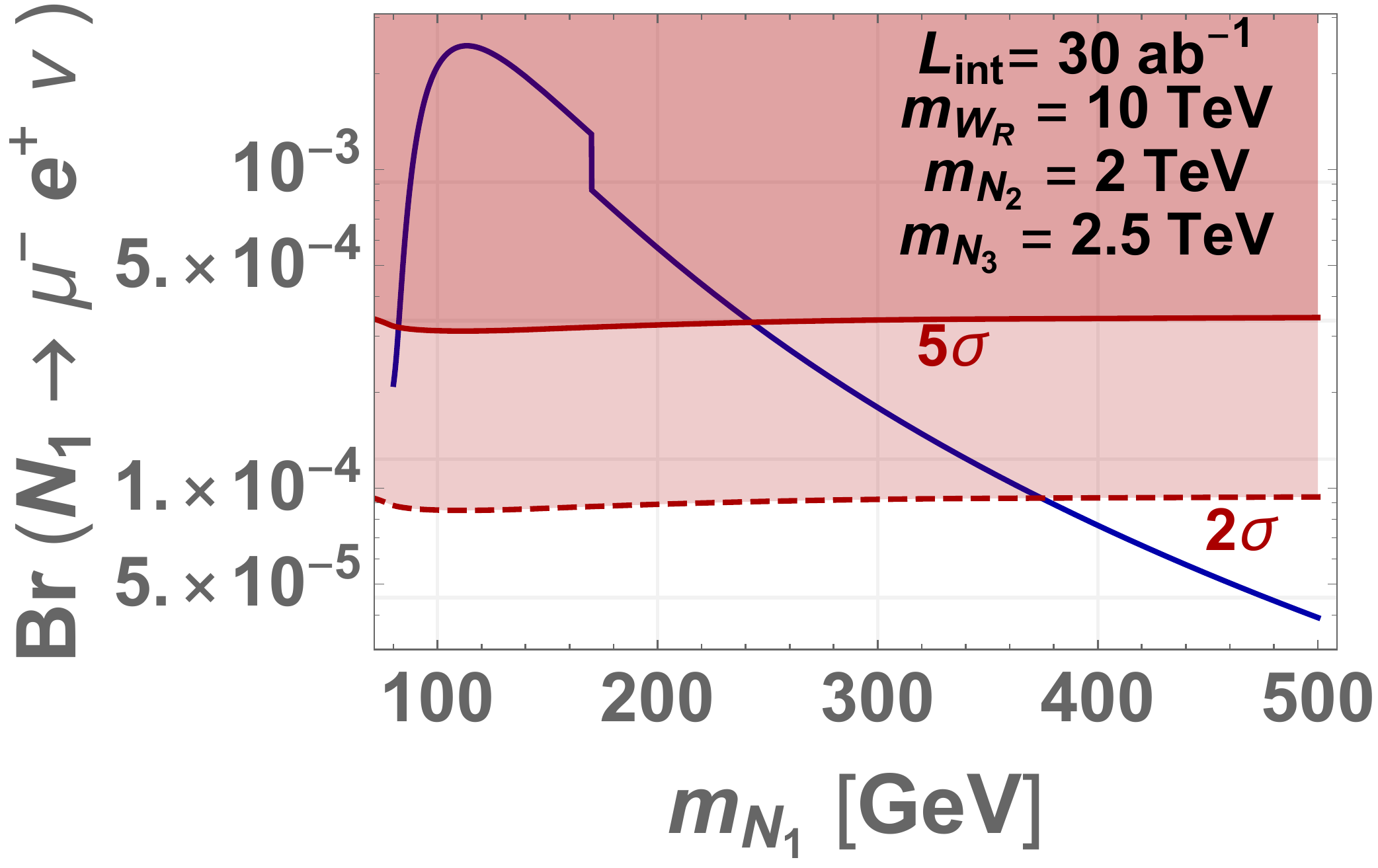}
      \includegraphics[scale=0.235]{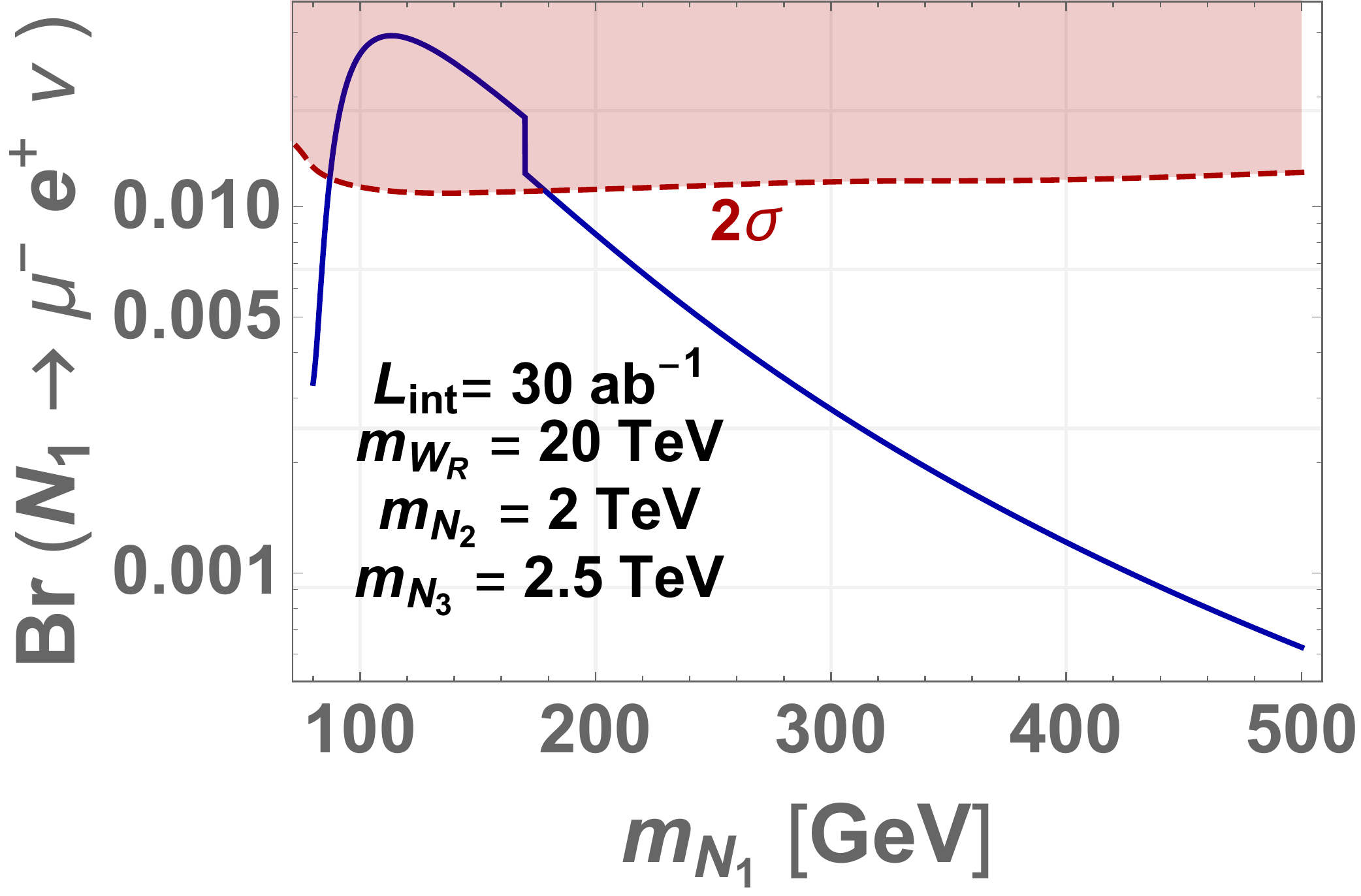}
  \caption{100 TeV $pp$ collider reach to the branching ratio of the purely leptonic decays of the heavy neutrino. The blue line denotes the branching ratio within the mLRSM and the shadowed thick(dashed) regions show the reach at $5\sigma(2\sigma)$, for an integrated luminosity of $L_{int} = 30 \text{ab}^{-1}$ and center of mass energy $\sqrt{s}= 100$ TeV. We assume $V_L=V_R^*$ and the  upper limit on light neutrino masses of $m_{\nu}=0.5$ eV~\cite{Nakamura:2010zzi}.     }
  \label{fig:br2lep2_FCC}
\end{figure}
\section{Connection between the Dirac mass,  the heavy and   light neutrino masses and low energy experiments}\label{discussion}

In this section, we discuss what the expected sensitivity obtained in the previous section implies for  the Dirac mass term of the minimal model. Summarizing,  we have the two following important points regarding the connection between the Dirac mass and the heavy and light neutrino masses: 
\begin{enumerate}
    \item The trilepton search can determine the mixing between the heavy and light neutrinos $\Theta_L$ in Eqs.~\eqref{theta_phys_P}  and \eqref{theta_phys_C} which, then, can  be related to the Dirac mass $M_D$ using Eq.~\eqref{HLmixing} once $M_N$ is known. The same conclusions apply for the non-minimal setup,  where the heavy light mixing is of the form $\Theta_L = \frac{1}{\sqrt{2}} M_D^{\dagger}  V_R \hat{M}_N^{-1}$ -- see Eq.~\eqref{mixing}. 
    \item The connection between $\Theta_L$,   $M_D$, $M_N$ and  $M_{\nu}$ is direct in the minimal model, as can be seen, for instance,  from Eq.~\eqref{ThetaHL_phys} for $\mathcal{C}$ as the LR symmetry. In the non-minimal model, relating experimentally accessible neutrino mass parameters with the Lagrangian mass parameters is far less direct due to the arbitrary, complex, orthogonal matrix $\mathcal{R}$ shown in Eq.~\eqref{sesaw}. 
\end{enumerate}

 \begin{figure}\centering
     \includegraphics[scale=0.357]{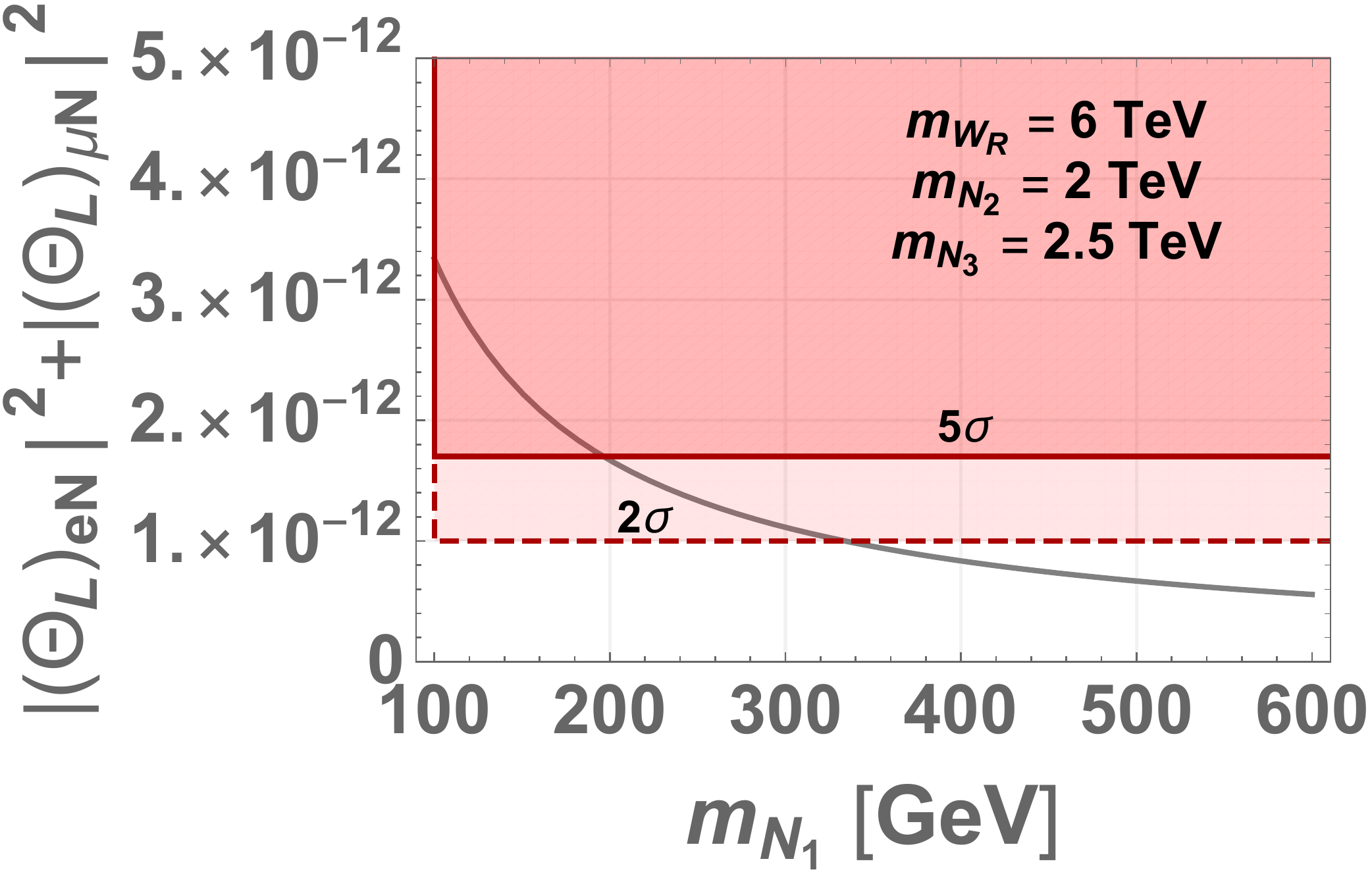}   \includegraphics[scale=0.31]{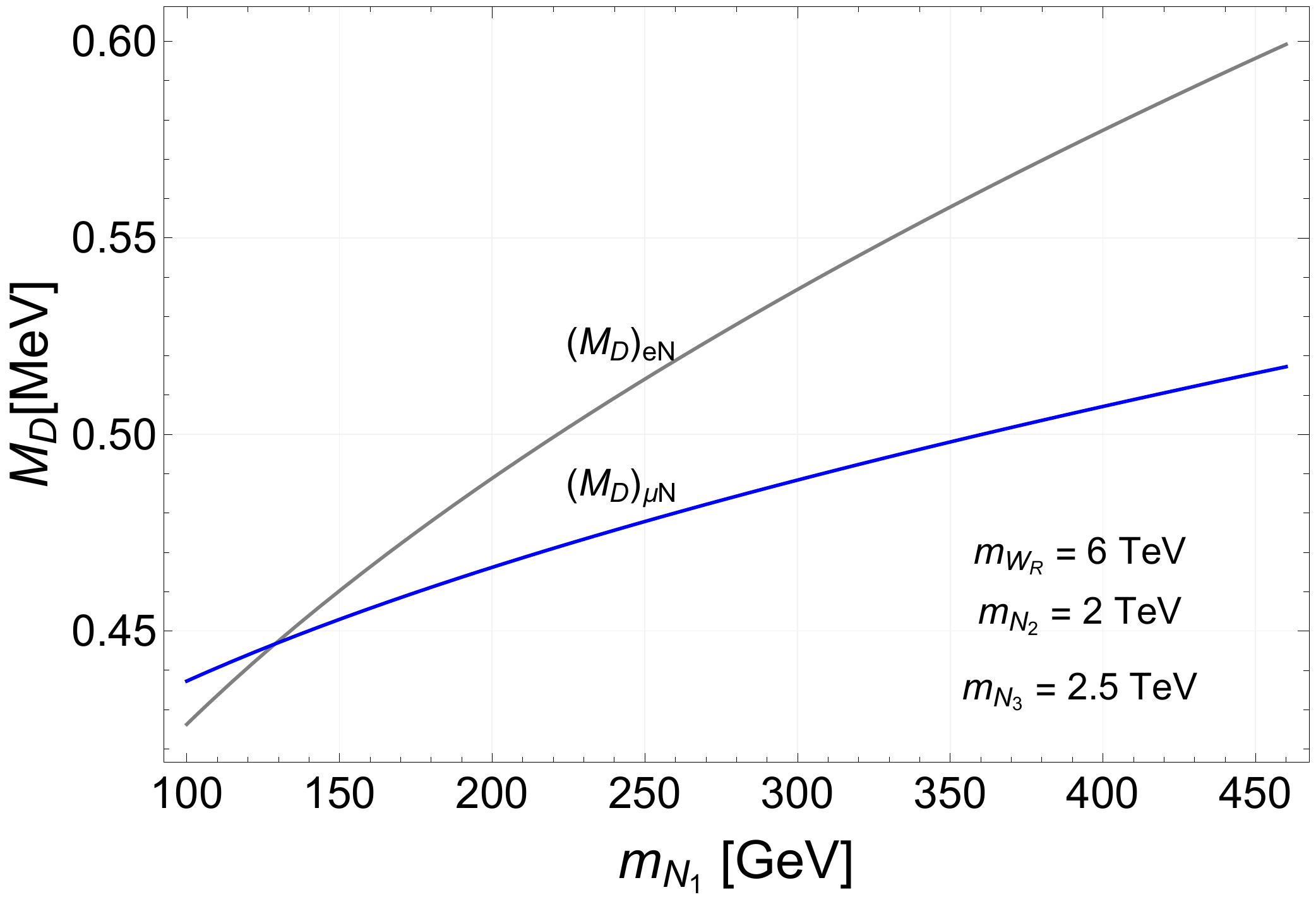}
  \caption{ Value for the heavy-light mixing angle combination  $|\Theta_{N\mu}|^2+|\Theta_{Ne}|^2$  (right)  and the Dirac Mass $M_D$ (left) as a function of the heavy neutrino mass, for $M_{W_R}= 6$ TeV   for a  $2\sigma$ exclusion region shown in Fig.~\ref{fig:br2lep2_FCC}. We assume $V_L=V_R^*$ and the  upper limit on light neutrino masses of $m_{\nu}=0.5$ eV~\cite{Nakamura:2010zzi}. }
  \label{thetaMD_FCC}
\end{figure}
 Therefore and for the mLRSM, it is interesting to compare the exclusion lines obtained in Fig.~\ref{fig:br2lep2_FCC} with the prediction one can make in this case.   In Fig.~\ref{thetaMD_FCC} (left panel)  we plot the combination $|(\Theta_L)_{N\mu}|^2+|(\Theta_L)_{Ne}|^2$ given by Eq.~\eqref{HLmixing}, as a function of $m_N$ together with the  $2\sigma$ and  $5\sigma$ significance regions expected at the 100 TeV $pp$ collider.  We see that at the 100 TeV $pp$ collider, the heavy-light mixing $\Theta$ can be probed for values as small as $|(\Theta_L)_{N\mu}|^2+|(\Theta_L)_{Ne}|^2 \sim 10^{-12}$. Notice that since $|(\Theta_L)_{N\mu}|^2+|(\Theta_L)_{Ne}|^2 \sim 10^{-12}$ is the sum of two positive terms one can safely assume that each  $|(\Theta_L)_{N\mu}|^2$ and $|(\Theta_L)_{Ne}|^2 \sim 10^{-12}$ are individually  smaller than $10^{-12}$.
 
 In what follows, we discuss how the above estimates translate to the sensitivity for the Dirac mass matrix elements $(M_D)_{eN}$ and $(M_D)_{\mu N}$ at the 100 TeV $pp$ collider.  To this end, it is instructive to show the relation between the Dirac mass matrix $M_D$ and the heavy, light neutrino mass matrices when $|(V_L)_{i,j}|=|(V_R)_{i,j}|$, since in this case the relation is simple enough to be written in a compact analytic form for both $\mathcal{C}$ and $\mathcal{P}$ cases. From Eq~\eqref{ThetaHL_phys} and  for $\mathcal{C}$ as the LR symmetry,  the Dirac  mass matrix $M_D$ can be written as~\cite{Nemevsek:2012iq}, 
  \begin{equation}\label{MD_C}
  M_D = V_L^* \hat{M}_N \sqrt{\frac{v_L}{v_R}-\frac{\hat{M}_{\nu}}{\hat{M}_N}}V_L^{\dagger}.
  \end{equation}
  Notice that this connection is lost for the non-minimal models, as can be explicitly seen in Eq.~\eqref{sesaw}, since in this case there is an orthogonal, complex matrix $\mathcal{R}$ which makes the Dirac mass arbitrary. 
  
  The same considerations apply also for $\mathcal{P}$ as the LR symmetry where the Dirac mass matrix can be written as\cite{Senjanovic:2016vxw}
   \begin{equation}\label{MD_P}
  M_D = V_L \hat{M}_N \sqrt{\frac{v_L}{v_R}-\frac{\hat{M}_{\nu}}{\hat{M}_N}}V_L^{\dagger}
  \end{equation}
These results explicitly show that when CP is an exact  symmetry, both LR symmetries coincide.  {When $|(V_L)_{i,j}|\neq |(V_R)_{i,j}|$,  the relation between the Dirac mass matrix and the heavy and light neutrino mass matrices shown in Eqs.~\eqref{MD_C} and \eqref{MD_P} is far more complicated, and this general situation has been discussed in Refs.~\cite{Nemevsek:2012iq,Senjanovic:2016vxw,Senjanovic:2018xtu}. Nevertheless, even  in this general  case,  the connection still exists and can be found by solving numerically some algebraic equations.}  

In short, the results we found apply for both Parity and Charge-conjugation as the LR symmetry.  In Fig.~\ref{thetaMD_FCC} (right panel),  we plot, using Eqs.~\eqref{MD_C} and \eqref{MD_P},  the values for $(M_D)_{eN}$ and $(M_D)_{\mu N}$ in the same range of heavy neutrino masses. We can conclude that Dirac mass terms of the order of $10^{-1}$ MeV can be probed at the 100 TeV $pp$ collider. { In the case the results from the  experiment does not match the  predictions made by the minimal model,  it would mean that  the discrete LR symmetry must be explicitly broken  in the Yukawa sector of the minimal model or that the LR symmetry is  non-minimally realized. } 

{{Finally, it is interesting to  observe that if approximate Parity symmetry is invoked as a solution to the Strong CP problem,  the neutron electric dipole moment~\cite{Ginges:2003qt,Pospelov:2005pr,Fukuyama:2012np,Engel:2013lsa} constraint, together with the constraint coming from the indirect CP violation ($\epsilon_K$) in the Kaon sector, would set $M_{W_R}\sim 20$ TeV~\cite{Maiezza:2014ala}. This scale is the same scale that may be probed at the $100$ TeV $pp$ collider. Hence,  the 100 TeV $pp$ collider offers the unique opportunity of finding the $W_R$ boson at the $20$ TeV scale, probing the Yukawa sector of the minimal model and offering a solution to the Strong CP problem by invoking Parity as an approximate symmetry at higher energies~\cite{Mohapatra:1978fy}. }}

\section{Conclusions}\label{conclusions}

{ In this work, we have analyzed the trilepton final state (produced via a $W_R$ gauge boson) as an effective channel for probing the Dirac mass terms of  neutrinos at hadronic colliders in the context of left-right symmetric models.  We have assessed the sensitivity to the heavy-light mixing of heavy neutrinos at the LHC, the HE-LHC and a 100 TeV $pp$ collider using the $p p \rightarrow W_R \rightarrow e^+ N \rightarrow e^+ (N \rightarrow e^+ W^-\rightarrow e^+ \mu^- \nu ) $ channel.   Within the minimal framework,  the relation between the Dirac mass matrix in terms of the heavy and light neutrino mass matrices implies that it is possible to translate the sensitivity to the heavy-light neutrino mixing  into a bound on the neutrino Dirac mass $M_D$.  For instance, we found that the minimal framework would not be seen at the LHC and the HE-LHC in the purely leptonic decays even with the ultimate integrated luminosities. Equivalently, this means that if any positive signal in the purely leptonic channel with the kinematic features described here is seen at the LHC, one would conclude left-right symmetry must be realized in a non-minimal context.

Finally, for a 100 TeV $pp$ collider with 30 $ab^{-1}$ of integrated luminosity,   we have found that for $W_R$ boson masses between $6-20$ TeV and heavy neutrinos masses between $80-460$ GeV, the Dirac mass term $(M_D)_{e N}$ and $(M_D)_{\mu N}$ can be excluded at $2\sigma$ up to masses of the order of $10^{-1}$ MeV and when either parity or charge conjugation is the left-right symmetry.  Furthermore, if the results of the experiment do not agree with  the  predictions  given by the minimal model, the  LR symmetry must be explicitly broken in the Yukawa sector or a non-minimal realization would be present.}

\section*{Acknowlegdements}
J.V. thanks Goran Senjanovi\'c for illuminating discussions and useful suggestions during all the stages of this work. J.C.H. thanks Martin Hirsch for useful discussions. J.V., J.C.H. and \mbox{N.N.} are grateful to the UMass Amherst Center for Fundamental Interactions for the hospitality, where they were visitors during the initial stages of this work.   J.V. was funded by FONDECYT grant No. 3170154 and by Conicyt PIA/Basal FB0821.  J.C.H. is supported by Chile grants Fondecyt No. 1161463 and Conicyt PIA/ACT 1406. MJRM and H-L.L was supported in part under U.S. Department of Energy Contract DE-SC0011095.
N.N. was supported by FONDECYT (Chile) grant 3170906.
H-L.L is also supported by the National Science Foundation of China under Grants No. 11875003

 %%%%%%%%%%%%%%%%%%%%%%%%%%%%%%%%%%%%%%%%%%%%%%%%%%%%%%%%%%%%%%%%%%%%%%%%%%%%%%%%%%%%%%%%%%%%%%%%%%%%%%%%%%%%%%%%%%%%%%%%%%%%%%%%%%%%%%%%%%%%%%%%%%%%%%%%%%%%%%%%%%%%%%%%%%%%%%%%%%

\bibliography{biblio}

\providecommand{\href}[2]{#2}\begingroup\raggedright\begin{thebibliography}{10}

\bibitem{Pati:1974yy}
J.~C. Pati and A.~Salam, \emph{{Lepton Number as the Fourth Color}},
  \href{http://dx.doi.org/10.1103/PhysRevD.10.275,
  10.1103/PhysRevD.11.703.2}{\emph{Phys. Rev.} {\bfseries D10} (1974)
  275--289}.

\bibitem{Mohapatra:1974gc}
R.~N. Mohapatra and J.~C. Pati, \emph{{A Natural Left-Right Symmetry}},
  \href{http://dx.doi.org/10.1103/PhysRevD.11.2558}{\emph{Phys. Rev.}
  {\bfseries D11} (1975) 2558}.

\bibitem{Senjanovic:1975rk}
G.~Senjanovi\'c and R.~N. Mohapatra, \emph{{Exact Left-Right Symmetry and
  Spontaneous Violation of Parity}},
  \href{http://dx.doi.org/10.1103/PhysRevD.12.1502}{\emph{Phys. Rev.}
  {\bfseries D12} (1975) 1502}.

\bibitem{Senjanovic:1978ev}
G.~Senjanovi\'c, \emph{{Spontaneous Breakdown of Parity in a Class of Gauge
  Theories}}, \href{http://dx.doi.org/10.1016/0550-3213(79)90604-7}{\emph{Nucl.
  Phys.} {\bfseries B153} (1979) 334--364}.

\bibitem{Mohapatra:1979ia}
R.~N. Mohapatra and G.~Senjanovi\'c, \emph{{Neutrino Mass and Spontaneous
  Parity Violation}},
  \href{http://dx.doi.org/10.1103/PhysRevLett.44.912}{\emph{Phys. Rev. Lett.}
  {\bfseries 44} (1980) 912}.

\bibitem{Mohapatra:1980yp}
R.~N. Mohapatra and G.~Senjanovi\'c, \emph{{Neutrino Masses and Mixings in
  Gauge Models with Spontaneous Parity Violation}},
  \href{http://dx.doi.org/10.1103/PhysRevD.23.165}{\emph{Phys. Rev.} {\bfseries
  D23} (1981) 165}.

\bibitem{Aad:2015zhl}
{\scshape ATLAS, CMS} collaboration, G.~Aad et~al., \emph{{Combined Measurement
  of the Higgs Boson Mass in $pp$ Collisions at $\sqrt{s}=7$ and 8 TeV with the
  ATLAS and CMS Experiments}},
  \href{http://dx.doi.org/10.1103/PhysRevLett.114.191803}{\emph{Phys. Rev.
  Lett.} {\bfseries 114} (2015) 191803},
  [\href{https://arxiv.org/abs/1503.07589}{{\ttfamily 1503.07589}}].

\bibitem{Chatrchyan:2014vua}
{\scshape CMS} collaboration, S.~Chatrchyan et~al., \emph{{Evidence for the
  direct decay of the 125 GeV Higgs boson to fermions}},
  \href{http://dx.doi.org/10.1038/nphys3005}{\emph{Nature Phys.} {\bfseries 10}
  (2014) 557--560}, [\href{https://arxiv.org/abs/1401.6527}{{\ttfamily
  1401.6527}}].

\bibitem{Minkowski:1977sc}
P.~Minkowski, \emph{{$\mu \to e\gamma$ at a Rate of One Out of $10^{9}$ Muon
  Decays?}}, \href{http://dx.doi.org/10.1016/0370-2693(77)90435-X}{\emph{Phys.
  Lett.} {\bfseries 67B} (1977) 421--428}.

\bibitem{Glashow:1979nm}
S.~L. Glashow, \emph{{The Future of Elementary Particle Physics}},
  \href{http://dx.doi.org/10.1007/978-1-4684-7197-7_15}{\emph{NATO Sci. Ser. B}
  {\bfseries 61} (1980) 687}.

\bibitem{GellMann:1980vs}
M.~Gell-Mann, P.~Ramond and R.~Slansky, \emph{{Complex Spinors and Unified
  Theories}}, {\emph{Conf. Proc.} {\bfseries C790927} (1979) 315--321},
  [\href{https://arxiv.org/abs/1306.4669}{{\ttfamily 1306.4669}}].

\bibitem{Yanagida:1979as}
T.~Yanagida, \emph{{HORIZONTAL SYMMETRY AND MASSES OF NEUTRINOS}}, {\emph{Conf.
  Proc.} {\bfseries C7902131} (1979) 95--99}.

\bibitem{Schechter:1980gr}
J.~Schechter and J.~W.~F. Valle, \emph{{Neutrino Masses in SU(2) x U(1)
  Theories}}, \href{http://dx.doi.org/10.1103/PhysRevD.22.2227}{\emph{Phys.
  Rev.} {\bfseries D22} (1980) 2227}.

\bibitem{Keung:1983uu}
W.-Y. Keung and G.~Senjanovi\'c, \emph{{Majorana Neutrinos and the Production
  of the Right-handed Charged Gauge Boson}},
  \href{http://dx.doi.org/10.1103/PhysRevLett.50.1427}{\emph{Phys. Rev. Lett.}
  {\bfseries 50} (1983) 1427}.

\bibitem{Nemevsek:2012iq}
M.~Nemevsek, G.~Senjanovi\'c and V.~Tello, \emph{{Connecting Dirac and Majorana
  Neutrino Mass Matrices in the Minimal Left-Right Symmetric Model}},
  \href{http://dx.doi.org/10.1103/PhysRevLett.110.151802}{\emph{Phys. Rev.
  Lett.} {\bfseries 110} (2013) 151802},
  [\href{https://arxiv.org/abs/1211.2837}{{\ttfamily 1211.2837}}].

\bibitem{Senjanovic:2016vxw}
G.~Senjanovi\'c and V.~Tello, \emph{{Probing Seesaw with Parity Restoration}},
  \href{http://dx.doi.org/10.1103/PhysRevLett.119.201803}{\emph{Phys. Rev.
  Lett.} {\bfseries 119} (2017) 201803},
  [\href{https://arxiv.org/abs/1612.05503}{{\ttfamily 1612.05503}}].

\bibitem{Senjanovic:2018xtu}
G.~Senjanovic and V.~Tello, \emph{{Disentangling Seesaw in the Minimal
  Left-Right Symmetric Model}},
  \href{https://arxiv.org/abs/1812.03790}{{\ttfamily 1812.03790}}.

\bibitem{Casas:2001sr}
J.~A. Casas and A.~Ibarra, \emph{{Oscillating neutrinos and muon ---> e,
  gamma}}, \href{http://dx.doi.org/10.1016/S0550-3213(01)00475-8}{\emph{Nucl.
  Phys.} {\bfseries B618} (2001) 171--204},
  [\href{https://arxiv.org/abs/hep-ph/0103065}{{\ttfamily hep-ph/0103065}}].

\bibitem{PhysRevD.62.013001}
A.~Ferrari, J.~Collot, M.-L. Andrieux, B.~Belhorma, P.~de~Saintignon, J.-Y.
  Hostachy et~al., \emph{Sensitivity study for new gauge bosons and
  right-handed majorana neutrinos in $\mathrm{pp}$ collisions at $\sqrt{s}=14
  \mathrm{TeV}$},
  \href{http://dx.doi.org/10.1103/PhysRevD.62.013001}{\emph{Phys. Rev. D}
  {\bfseries 62} (May, 2000) 013001}.

\bibitem{Han:2012vk}
T.~Han, I.~Lewis, R.~Ruiz and Z.-g. Si, \emph{{Lepton Number Violation and
  $W^\prime$ Chiral Couplings at the LHC}},
  \href{http://dx.doi.org/10.1103/PhysRevD.87.035011,
  10.1103/PhysRevD.87.039906}{\emph{Phys. Rev.} {\bfseries D87} (2013) 035011},
  [\href{https://arxiv.org/abs/1211.6447}{{\ttfamily 1211.6447}}].

\bibitem{Izaguirre:2015pga}
E.~Izaguirre and B.~Shuve, \emph{{Multilepton and Lepton Jet Probes of
  Sub-Weak-Scale Right-Handed Neutrinos}},
  \href{http://dx.doi.org/10.1103/PhysRevD.91.093010}{\emph{Phys. Rev.}
  {\bfseries D91} (2015) 093010},
  [\href{https://arxiv.org/abs/1504.02470}{{\ttfamily 1504.02470}}].

\bibitem{Sirunyan:2018mtv}
{\scshape CMS} collaboration, A.~M. Sirunyan et~al., \emph{{Search for heavy
  neutral leptons in events with three charged leptons in proton-proton
  collisions at $\sqrt{s} =$ 13 TeV}},
  \href{http://dx.doi.org/10.1103/PhysRevLett.120.221801}{\emph{Phys. Rev.
  Lett.} {\bfseries 120} (2018) 221801},
  [\href{https://arxiv.org/abs/1802.02965}{{\ttfamily 1802.02965}}].

\bibitem{Das:2016akd}
A.~Das, N.~Nagata and N.~Okada, \emph{{Testing the 2-TeV Resonance with
  Trileptons}}, \href{http://dx.doi.org/10.1007/JHEP03(2016)049}{\emph{JHEP}
  {\bfseries 03} (2016) 049},
  [\href{https://arxiv.org/abs/1601.05079}{{\ttfamily 1601.05079}}].

\bibitem{Kersten:2007vk}
J.~Kersten and A.~{\relax Yu}. Smirnov, \emph{{Right-Handed Neutrinos at CERN
  LHC and the Mechanism of Neutrino Mass Generation}},
  \href{http://dx.doi.org/10.1103/PhysRevD.76.073005}{\emph{Phys. Rev.}
  {\bfseries D76} (2007) 073005},
  [\href{https://arxiv.org/abs/0705.3221}{{\ttfamily 0705.3221}}].

\bibitem{deGouvea:2007hks}
A.~de~Gouvea, \emph{{GeV seesaw, accidentally small neutrino masses, and Higgs
  decays to neutrinos}},  \href{https://arxiv.org/abs/0706.1732}{{\ttfamily
  0706.1732}}.

\bibitem{Xing:2009in}
Z.-z. Xing, \emph{{Naturalness and Testability of TeV Seesaw Mechanisms}},
  \href{http://dx.doi.org/10.1143/PTPS.180.112}{\emph{Prog. Theor. Phys.
  Suppl.} {\bfseries 180} (2009) 112--127},
  [\href{https://arxiv.org/abs/0905.3903}{{\ttfamily 0905.3903}}].

\bibitem{He:2009ua}
X.-G. He, S.~Oh, J.~Tandean and C.-C. Wen, \emph{{Large Mixing of Light and
  Heavy Neutrinos in Seesaw Models and the LHC}},
  \href{http://dx.doi.org/10.1103/PhysRevD.80.073012}{\emph{Phys. Rev.}
  {\bfseries D80} (2009) 073012},
  [\href{https://arxiv.org/abs/0907.1607}{{\ttfamily 0907.1607}}].

\bibitem{Ibarra:2010xw}
A.~Ibarra, E.~Molinaro and S.~T. Petcov, \emph{{TeV Scale See-Saw Mechanisms of
  Neutrino Mass Generation, the Majorana Nature of the Heavy Singlet Neutrinos
  and $(\beta\beta)_{0\nu}$-Decay}},
  \href{http://dx.doi.org/10.1007/JHEP09(2010)108}{\emph{JHEP} {\bfseries 09}
  (2010) 108}, [\href{https://arxiv.org/abs/1007.2378}{{\ttfamily 1007.2378}}].

\bibitem{Haba:2011pe}
N.~Haba, T.~Horita, K.~Kaneta and Y.~Mimura, \emph{{TeV-scale seesaw with
  non-negligible left-right neutrino mixings}},
  \href{https://arxiv.org/abs/1110.2252}{{\ttfamily 1110.2252}}.

\bibitem{Mitra:2011qr}
M.~Mitra, G.~Senjanovic and F.~Vissani, \emph{{Neutrinoless Double Beta Decay
  and Heavy Sterile Neutrinos}},
  \href{http://dx.doi.org/10.1016/j.nuclphysb.2011.10.035}{\emph{Nucl. Phys.}
  {\bfseries B856} (2012) 26--73},
  [\href{https://arxiv.org/abs/1108.0004}{{\ttfamily 1108.0004}}].

\bibitem{Chen:2013fna}
C.-Y. Chen, P.~S.~B. Dev and R.~N. Mohapatra, \emph{{Probing Heavy-Light
  Neutrino Mixing in Left-Right Seesaw Models at the LHC}},
  \href{http://dx.doi.org/10.1103/PhysRevD.88.033014}{\emph{Phys. Rev.}
  {\bfseries D88} (2013) 033014},
  [\href{https://arxiv.org/abs/1306.2342}{{\ttfamily 1306.2342}}].

\bibitem{Akhmedov:2006de}
E.~K. Akhmedov and M.~Frigerio, \emph{{Interplay of type I and type II seesaw
  contributions to neutrino mass}},
  \href{http://dx.doi.org/10.1088/1126-6708/2007/01/043}{\emph{JHEP} {\bfseries
  01} (2007) 043}, [\href{https://arxiv.org/abs/hep-ph/0609046}{{\ttfamily
  hep-ph/0609046}}].

\bibitem{Akhmedov:2006yp}
E.~K. Akhmedov, M.~Blennow, T.~Hallgren, T.~Konstandin and T.~Ohlsson,
  \emph{{Stability and leptogenesis in the left-right symmetric seesaw
  mechanism}},
  \href{http://dx.doi.org/10.1088/1126-6708/2007/04/022}{\emph{JHEP} {\bfseries
  04} (2007) 022}, [\href{https://arxiv.org/abs/hep-ph/0612194}{{\ttfamily
  hep-ph/0612194}}].

\bibitem{Chao:2007mz}
W.~Chao, S.~Luo, Z.-z. Xing and S.~Zhou, \emph{{A Compromise between Neutrino
  Masses and Collider Signatures in the Type-II Seesaw Model}},
  \href{http://dx.doi.org/10.1103/PhysRevD.77.016001}{\emph{Phys. Rev.}
  {\bfseries D77} (2008) 016001},
  [\href{https://arxiv.org/abs/0709.1069}{{\ttfamily 0709.1069}}].

\bibitem{Senjanovic:2014pva}
G.~Senjanović and V.~Tello, \emph{{Right Handed Quark Mixing in Left-Right
  Symmetric Theory}},
  \href{http://dx.doi.org/10.1103/PhysRevLett.114.071801}{\emph{Phys. Rev.
  Lett.} {\bfseries 114} (2015) 071801},
  [\href{https://arxiv.org/abs/1408.3835}{{\ttfamily 1408.3835}}].

\bibitem{Senjanovic:2015yea}
G.~Senjanović and V.~Tello, \emph{{Restoration of Parity and the Right-Handed
  Analog of the CKM Matrix}},
  \href{http://dx.doi.org/10.1103/PhysRevD.94.095023}{\emph{Phys. Rev.}
  {\bfseries D94} (2016) 095023},
  [\href{https://arxiv.org/abs/1502.05704}{{\ttfamily 1502.05704}}].

\bibitem{Dube:2017jgo}
S.~Dube, D.~Gadkari and A.~M. Thalapillil, \emph{{Lepton-Jets and Low-Mass
  Sterile Neutrinos at Hadron Colliders}},
  \href{http://dx.doi.org/10.1103/PhysRevD.96.055031}{\emph{Phys. Rev.}
  {\bfseries D96} (2017) 055031},
  [\href{https://arxiv.org/abs/1707.00008}{{\ttfamily 1707.00008}}].

\bibitem{Tello:2012qda}
V.~Tello, \emph{{Connections between the high and low energy violation of
  Lepton and Flavor numbers in the minimal left-right symmetric model}}.
\newblock PhD thesis, SISSA, Trieste, 2012.

\bibitem{Maiezza:2016ybz}
A.~Maiezza, G.~Senjanovi\'c and J.~C. Vasquez, \emph{{Higgs sector of the
  minimal left-right symmetric theory}},
  \href{http://dx.doi.org/10.1103/PhysRevD.95.095004}{\emph{Phys. Rev.}
  {\bfseries D95} (2017) 095004},
  [\href{https://arxiv.org/abs/1612.09146}{{\ttfamily 1612.09146}}].

\bibitem{Parida:2010wq}
M.~K. Parida and A.~Raychaudhuri, \emph{{Inverse see-saw, leptogenesis,
  observable proton decay and $\Delta^{\pm\pm}_{\rm R}$ in SUSY SO(10) with
  heavy $W_R$}},
  \href{http://dx.doi.org/10.1103/PhysRevD.82.093017}{\emph{Phys. Rev.}
  {\bfseries D82} (2010) 093017},
  [\href{https://arxiv.org/abs/1007.5085}{{\ttfamily 1007.5085}}].

\bibitem{Arbelaez:2013nga}
C.~Arbeláez, M.~Hirsch, M.~Malinský and J.~C. Romão, \emph{{LHC-scale
  left-right symmetry and unification}},
  \href{http://dx.doi.org/10.1103/PhysRevD.89.035002}{\emph{Phys. Rev.}
  {\bfseries D89} (2014) 035002},
  [\href{https://arxiv.org/abs/1311.3228}{{\ttfamily 1311.3228}}].

\bibitem{Mohapatra:1986bd}
R.~N. Mohapatra and J.~W.~F. Valle, \emph{{Neutrino Mass and Baryon Number
  Nonconservation in Superstring Models}},
  \href{http://dx.doi.org/10.1103/PhysRevD.34.1642}{\emph{Phys. Rev.}
  {\bfseries D34} (1986) 1642}.

\bibitem{Dev:2015pga}
P.~S. Bhupal~Dev and R.~N. Mohapatra, \emph{{Unified explanation of the $eejj$,
  diboson and dijet resonances at the LHC}},
  \href{http://dx.doi.org/10.1103/PhysRevLett.115.181803}{\emph{Phys. Rev.
  Lett.} {\bfseries 115} (2015) 181803},
  [\href{https://arxiv.org/abs/1508.02277}{{\ttfamily 1508.02277}}].

\bibitem{Anamiati:2016uxp}
G.~Anamiati, M.~Hirsch and E.~Nardi, \emph{{Quasi-Dirac neutrinos at the LHC}},
  \href{http://dx.doi.org/10.1007/JHEP10(2016)010}{\emph{JHEP} {\bfseries 10}
  (2016) 010}, [\href{https://arxiv.org/abs/1607.05641}{{\ttfamily
  1607.05641}}].

\bibitem{Anamiati:2017rxw}
G.~Anamiati, R.~M. Fonseca and M.~Hirsch, \emph{{Quasi Dirac neutrino
  oscillations}},
  \href{http://dx.doi.org/10.1103/PhysRevD.97.095008}{\emph{Phys. Rev.}
  {\bfseries D97} (2018) 095008},
  [\href{https://arxiv.org/abs/1710.06249}{{\ttfamily 1710.06249}}].

\bibitem{Mitra:2016kov}
M.~Mitra, R.~Ruiz, D.~J. Scott and M.~Spannowsky, \emph{{Neutrino Jets from
  High-Mass $W_R$ Gauge Bosons in TeV-Scale Left-Right Symmetric Models}},
  \href{http://dx.doi.org/10.1103/PhysRevD.94.095016}{\emph{Phys. Rev.}
  {\bfseries D94} (2016) 095016},
  [\href{https://arxiv.org/abs/1607.03504}{{\ttfamily 1607.03504}}].

\bibitem{Nemevsek:2018bbt}
M.~Nemevsek, F.~Nesti and G.~Popara, \emph{{Keung-Senjanovi\'c process at LHC:
  from LNV to displaced vertices to invisible decays}},
  \href{https://arxiv.org/abs/1801.05813}{{\ttfamily 1801.05813}}.

\bibitem{Nakamura:2010zzi}
{\scshape Particle Data Group} collaboration, K.~Nakamura et~al., \emph{{Review
  of particle physics}},
  \href{http://dx.doi.org/10.1088/0954-3899/37/7A/075021}{\emph{J. Phys.}
  {\bfseries G37} (2010) 075021}.

\bibitem{Arbelaez:2017zqq}
C.~Arbela�z, C.~Dib, I.~Schmidt and J.~C. Vasquez, \emph{{Probing the Dirac
  or Majorana nature of the Heavy Neutrinos in pure leptonic decays at the
  LHC}}, \href{http://dx.doi.org/10.1103/PhysRevD.97.055011}{\emph{Phys. Rev.}
  {\bfseries D97} (2018) 055011},
  [\href{https://arxiv.org/abs/1712.08704}{{\ttfamily 1712.08704}}].

\bibitem{Alloul:2013bka}
A.~Alloul, N.~D. Christensen, C.~Degrande, C.~Duhr and B.~Fuks,
  \emph{{FeynRules 2.0 - A complete toolbox for tree-level phenomenology}},
  \href{http://dx.doi.org/10.1016/j.cpc.2014.04.012}{\emph{Comput. Phys.
  Commun.} {\bfseries 185} (2014) 2250--2300},
  [\href{https://arxiv.org/abs/1310.1921}{{\ttfamily 1310.1921}}].

\bibitem{Roitgrund:2014zka}
A.~Roitgrund, G.~Eilam and S.~Bar-Shalom, \emph{{Implementation of the
  left-right symmetric model in FeynRules}},
  \href{http://dx.doi.org/10.1016/j.cpc.2015.12.009}{\emph{Comput. Phys.
  Commun.} {\bfseries 203} (2016) 18--44},
  [\href{https://arxiv.org/abs/1401.3345}{{\ttfamily 1401.3345}}].

\bibitem{Nemevsek:2016enw}
M.~Nemevsek, F.~Nesti and J.~C. Vasquez, \emph{{Majorana Higgses at
  colliders}}, \href{http://dx.doi.org/10.1007/JHEP04(2017)114}{\emph{JHEP}
  {\bfseries 04} (2017) 114},
  [\href{https://arxiv.org/abs/1612.06840}{{\ttfamily 1612.06840}}].

\bibitem{Alwall:2014hca}
J.~Alwall, R.~Frederix, S.~Frixione, V.~Hirschi, F.~Maltoni, O.~Mattelaer
  et~al., \emph{{The automated computation of tree-level and next-to-leading
  order differential cross sections, and their matching to parton shower
  simulations}}, \href{http://dx.doi.org/10.1007/JHEP07(2014)079}{\emph{JHEP}
  {\bfseries 07} (2014) 079},
  [\href{https://arxiv.org/abs/1405.0301}{{\ttfamily 1405.0301}}].

\bibitem{Sjostrand:2006za}
T.~Sjostrand, S.~Mrenna and P.~Z. Skands, \emph{{PYTHIA 6.4 Physics and
  Manual}}, \href{http://dx.doi.org/10.1088/1126-6708/2006/05/026}{\emph{JHEP}
  {\bfseries 05} (2006) 026},
  [\href{https://arxiv.org/abs/hep-ph/0603175}{{\ttfamily hep-ph/0603175}}].

\bibitem{deFavereau:2013fsa}
{\scshape DELPHES 3} collaboration, J.~de~Favereau, C.~Delaere, P.~Demin,
  A.~Giammanco, V.~Lema�tre, A.~Mertens et~al., \emph{{DELPHES 3, A modular
  framework for fast simulation of a generic collider experiment}},
  \href{http://dx.doi.org/10.1007/JHEP02(2014)057}{\emph{JHEP} {\bfseries 02}
  (2014) 057}, [\href{https://arxiv.org/abs/1307.6346}{{\ttfamily 1307.6346}}].

\bibitem{Aaboud:2017qph}
{\scshape ATLAS} collaboration, M.~Aaboud et~al., \emph{{Search for doubly
  charged Higgs boson production in multi-lepton final states with the ATLAS
  detector using proton–proton collisions at $\sqrt{s}=13\,\text {TeV}$}},
  \href{http://dx.doi.org/10.1140/EPJC/S10052-018-5661-Z,
  10.1140/epjc/s10052-018-5661-z}{\emph{Eur. Phys. J.} {\bfseries C78} (2018)
  199}, [\href{https://arxiv.org/abs/1710.09748}{{\ttfamily 1710.09748}}].

\bibitem{Nemevsek:2011hz}
M.~Nemevsek, F.~Nesti, G.~Senjanovi\'c and Y.~Zhang, \emph{{First Limits on
  Left-Right Symmetry Scale from LHC Data}},
  \href{http://dx.doi.org/10.1103/PhysRevD.83.115014}{\emph{Phys. Rev.}
  {\bfseries D83} (2011) 115014},
  [\href{https://arxiv.org/abs/1103.1627}{{\ttfamily 1103.1627}}].

\bibitem{Helo:2013esa}
J.~C. Helo, M.~Hirsch and S.~Kovalenko, \emph{{Heavy neutrino searches at the
  LHC with displaced vertices}},
  \href{http://dx.doi.org/10.1103/PhysRevD.89.073005,
  10.1103/PhysRevD.93.099902}{\emph{Phys. Rev.} {\bfseries D89} (2014) 073005},
  [\href{https://arxiv.org/abs/1312.2900}{{\ttfamily 1312.2900}}].

\bibitem{Cottin:2018kmq}
G.~Cottin, J.~C. Helo and M.~Hirsch, \emph{{Searches for light sterile
  neutrinos with multitrack displaced vertices}},
  \href{http://dx.doi.org/10.1103/PhysRevD.97.055025}{\emph{Phys. Rev.}
  {\bfseries D97} (2018) 055025},
  [\href{https://arxiv.org/abs/1801.02734}{{\ttfamily 1801.02734}}].

\bibitem{Cottin:2018nms}
G.~Cottin, J.~C. Helo and M.~Hirsch, \emph{{Displaced vertices as probes of
  sterile neutrino mixing at the LHC}},
  \href{http://dx.doi.org/10.1103/PhysRevD.98.035012}{\emph{Phys. Rev.}
  {\bfseries D98} (2018) 035012},
  [\href{https://arxiv.org/abs/1806.05191}{{\ttfamily 1806.05191}}].

\bibitem{Das:2017hmg}
A.~Das, P.~S.~B. Dev and R.~N. Mohapatra, \emph{{Same Sign versus Opposite Sign
  Dileptons as a Probe of Low Scale Seesaw Mechanisms}},
  \href{http://dx.doi.org/10.1103/PhysRevD.97.015018}{\emph{Phys. Rev.}
  {\bfseries D97} (2018) 015018},
  [\href{https://arxiv.org/abs/1709.06553}{{\ttfamily 1709.06553}}].

\bibitem{Mangano:2016jyj}
M.~L. Mangano et~al., \emph{{Physics at a 100 TeV pp Collider: Standard Model
  Processes}}, \href{http://dx.doi.org/10.23731/CYRM-2017-003.1}{\emph{CERN
  Yellow Report} (2017) 1--254},
  [\href{https://arxiv.org/abs/1607.01831}{{\ttfamily 1607.01831}}].

\bibitem{Ginges:2003qt}
J.~S.~M. Ginges and V.~V. Flambaum, \emph{{Violations of fundamental symmetries
  in atoms and tests of unification theories of elementary particles}},
  \href{http://dx.doi.org/10.1016/j.physrep.2004.03.005}{\emph{Phys. Rept.}
  {\bfseries 397} (2004) 63--154},
  [\href{https://arxiv.org/abs/physics/0309054}{{\ttfamily physics/0309054}}].

\bibitem{Pospelov:2005pr}
M.~Pospelov and A.~Ritz, \emph{{Electric dipole moments as probes of new
  physics}}, \href{http://dx.doi.org/10.1016/j.aop.2005.04.002}{\emph{Annals
  Phys.} {\bfseries 318} (2005) 119--169},
  [\href{https://arxiv.org/abs/hep-ph/0504231}{{\ttfamily hep-ph/0504231}}].

\bibitem{Fukuyama:2012np}
T.~Fukuyama, \emph{{Searching for New Physics beyond the Standard Model in
  Electric Dipole Moment}},
  \href{http://dx.doi.org/10.1142/S0217751X12300153}{\emph{Int. J. Mod. Phys.}
  {\bfseries A27} (2012) 1230015},
  [\href{https://arxiv.org/abs/1201.4252}{{\ttfamily 1201.4252}}].

\bibitem{Engel:2013lsa}
J.~Engel, M.~J. Ramsey-Musolf and U.~van Kolck, \emph{{Electric Dipole Moments
  of Nucleons, Nuclei, and Atoms: The Standard Model and Beyond}},
  \href{http://dx.doi.org/10.1016/j.ppnp.2013.03.003}{\emph{Prog. Part. Nucl.
  Phys.} {\bfseries 71} (2013) 21--74},
  [\href{https://arxiv.org/abs/1303.2371}{{\ttfamily 1303.2371}}].

\bibitem{Maiezza:2014ala}
A.~Maiezza and M.~Nemevsek, \emph{{Strong P invariance, neutron electric dipole
  moment, and minimal left-right parity at LHC}},
  \href{http://dx.doi.org/10.1103/PhysRevD.90.095002}{\emph{Phys. Rev.}
  {\bfseries D90} (2014) 095002},
  [\href{https://arxiv.org/abs/1407.3678}{{\ttfamily 1407.3678}}].

\bibitem{Mohapatra:1978fy}
R.~N. Mohapatra and G.~Senjanovi\'c, \emph{{Natural Suppression of Strong p and
  t Noninvariance}},
  \href{http://dx.doi.org/10.1016/0370-2693(78)90243-5}{\emph{Phys. Lett.}
  {\bfseries 79B} (1978) 283--286}.

\end{thebibliography}\endgroup

\bibliographystyle{jhep}

\end{document}